\documentclass[10pt,a4paper,twocolumn]{IEEEtran}
\usepackage{amsopn,amsmath,amssymb,amsfonts,bm,amsthm}
\usepackage{algorithm,algorithmic}
\usepackage{mdwlist}
\usepackage{flushend,cuted}
\usepackage{graphicx,subfigure}
\usepackage{enumerate}
\usepackage{cite}
\usepackage{etex}
\usepackage{tikz,pgfplots}
\usepackage{array,multirow}
\usetikzlibrary{plotmarks}
\pgfplotsset{compat=1.5}
\newcommand{\set}[1]{\mathcal{#1}}

\def\C{\set{C}}
\def\E{\set{E}}
\def\f{\mathcal F}
\def\F{\mathbb{F}}
\def\G{\set{G}}
\def\K{\set{K}}
\def\M{\set{M}}

\def\P{\set{P}}
\def\p{\mathbf p}
\def\V{\set{V}}
\def\X{\boldsymbol X}
\def\S{\set{S}}
\def\cG{\overline{\set{G}}}
\def\cGn{\overline{\set{G'}}}

\def\mA{\boldsymbol A}
\def\mC{\boldsymbol C}

\def\UIDNC{U_{\textrm{IDNC}}}

\def\DIDNC{D_{\textrm{IDNC}}}

\def\w{\omega}

\def\eqv{\Leftrightarrow}

\def\conf{\oslash}

\def\T{\mathcal T}
\def\R{\mathcal R}

\def\W{\set{W}}

\newcommand{\secref}[1]{Section\,\ref{#1}}

\newcommand{\figref}[1]{Fig.\,\ref{#1}}

\newtheorem{Lemma}{\textbf{Lemma}}
\newtheorem{Theorem}{\textbf{Theorem}}

\newtheorem{Definition}{\textbf{Definition}}

\newtheorem{Remark}{\textbf{Remark}}

\newtheorem{Example}{\textbf{Example}}
  {\proof}{\proofend}

\author{
  \IEEEauthorblockA{Mingchao Yu~~~~~~~~~~~~~~~~Parastoo Sadeghi~~~~~~~~~~~~~~~~Neda Aboutorab}\\
  \IEEEauthorblockA{\small Research School of Engineering, The Australian National University, Canberra, Australia \\ \texttt{\{ming.yu, parastoo.sadeghi, neda.aboutorab\}@anu.edu.au}}}
\title{{On Throughput and Decoding Delay Performance of \\ Instantly Decodable Network Coding}}
\begin{document}
\sloppy

\maketitle

\begin{abstract}
In this paper, a comprehensive study of packet-based instantly decodable network coding (IDNC) for single-hop wireless broadcast is presented. The optimal IDNC solution in terms of throughput is proposed and its packet decoding delay performance is investigated. Lower and upper bounds on the achievable throughput and decoding delay performance of IDNC are derived and assessed through extensive simulations. Furthermore, the impact of receivers' feedback frequency on the performance of IDNC is studied and optimal IDNC solutions are proposed for scenarios where receivers' feedback is only available after an IDNC round, composed of several coded transmissions.  However, since finding these IDNC optimal solutions is computationally complex, we further propose simple yet efficient heuristic IDNC algorithms. The impact of system settings and parameters such as channel erasure probability, feedback frequency, and the number of receivers is also investigated and simple guidelines for practical implementations of IDNC are proposed.
\end{abstract}

\section{Introduction}
Instantly decodable network coding (IDNC) is a class of linear network coding schemes \cite{katti1;etal:2008,keller:drinea:fragouli:2008,costa:munaretto:widmer:baros:2008,Rozner_Heuristic_clique,sadeghi:shams:traskov:2010,sorour:valaee:2011,le:dimakis:idnc_real} which has been widely studied and applied in wireless unicast, multicast and broadcast systems. It has been shown that IDNC schemes can quite significantly improve data throughput in such systems compared to their uncoded counterparts \cite{katti1;etal:2008,keller:drinea:fragouli:2008}, while offering simple XOR-based encoding and decoding. Furthermore, IDNC provides instant packet decodability at the receivers, which can result in faster delivery of the packets to the application layer compared to other linear network coding schemes \cite{sadeghi:shams:traskov:2010,sorour:valaee:2011}.

In this paper, we are primarily concerned with investigating the performance limits of IDNC schemes for data dissemination in single-hop wireless broadcast systems in terms of throughput and delay. In such systems, there is a single sender who wishes to broadcast a block of data packets to multiple receivers \cite{Rozner_Heuristic_clique,sadeghi:shams:traskov:2010,sorour:valaee:2011,le:dimakis:idnc_real}. Due to packet erasures in wireless fading channels, some transmitted packets are lost at the receivers. Generally, information about received or lost packets are fed back from the receivers to the sender after transmission of one or multiple packets. The sender then determines which data packets from the block to combine and transmit next subject to IDNC constraints. This process is repeated until the broadcast of the block is complete, i.e. until all receivers have decoded all data packets.

In such systems, the time it takes to complete the block, or simply the IDNC block completion time, is a fundamental measure of its throughput performance and will be studied in this paper. Taking the block completion time of random linear network coding (RLNC) \cite{ho:medard:koetter:karger:effros:2006} as a benchmark, many works in the literature have been concerned with proposing IDNC schemes with good throughput performance \cite{Rozner_Heuristic_clique,sadeghi:shams:traskov:2010,sorour:valaee:2011,sameh:valaee:globecom:2010,
sadeghi:adaptive_broadcast_2009,sorour:limited_feedback:2011,sorour:lossy_feedback:2011,wang:sensor:2010,li:idnc_video:2011,karim:memory:2012,le:dimakis:idnc_real}. The majority of these schemes collect feedback about the lost packets and determine an \emph{online} IDNC solution accordingly, which comprises one or more coded packets, such that it can efficiently bring the system closer to block completion.

Although these works differ in their models and assumptions about the frequency and reliability of feedback \cite{sorour:limited_feedback:2011,sorour:lossy_feedback:2011}, at their core they run dynamic IDNC algorithms, responding to erasure patterns that have happened along the transmission. The main limitation of such studies is that it is impossible to say \emph{a priori} how long it will actually take to complete a block starting with a certain system packet reception state at the receivers. Furthermore, such IDNC solution is in fact the result of a local optimization, and there is no guarantee that the solution is globally optimal. Therefore, the following two fundamental questions still seem unanswered in the literature:
\begin{enumerate}
\item What is the \emph{best} throughput performance of IDNC?
\item Which IDNC solution can achieve this best throughput performance?
\end{enumerate}
By best throughput, we refer to the minimum block completion time that is possible by using IDNC starting with a certain system packet reception state, \emph{in the absence of any future erasures}. It is clear that packet erasures in an actual system can defer block completion. Therefore, our measure of throughput is the best possible performance of the IDNC schemes in terms of throughput and serves as an upper bound on what IDNC can achieve in the presence of erasures. Such measure of throughput is significant because it disentangles the effects of channel-induced packet erasures and algorithm-induced IDNC coded packet selection on the throughput of the system. Based on this measure of throughput, we propose the concept of \emph{optimal} IDNC scheme which refers to an IDNC scheme that provides globally optimal IDNC solution and achieves the best throughput performance in the absence of any future erasures.

Besides block completion time, another fundamental performance metric of IDNC is its decoding delay. There are various definitions of decoding delay in the literature \cite{katti1;etal:2008,costa:munaretto:widmer:baros:2008,sadeghi:shams:traskov:2010}. In this work, we consider \emph{packet decoding delay}, defined as the number of time slots it takes till the data packets are decoded by the receivers. The reasons for our choice are twofold. First, short packet decoding delay is the main advantage of IDNC, which is particularly desirable for applications in which data packets are useful regardless of their order. Second, packet decoding delay is naturally related to the throughput of IDNC and its respective IDNC solution. That is, having investigated the best throughput performance of IDNC and the optimal IDNC solution, decoding delay limits of IDNC schemes can be obtained with relative ease.

The contributions of this paper can be summarized as follows.
First, the best achievable throughput performance of IDNC, regardless of packets erasure probabilities and feedback frequency, and its corresponding optimal IDNC solution are rigorously obtained. Furthermore, the concept of \emph{IDNC packet diversity} in the optimal IDNC solution is introduced. It is a measure of the robustness of IDNC solution against packet erasures. While ensuring the optimal throughput performance, our proposed IDNC solution enhances packet diversity wherever possible, hence enhancing its robustness against erasures. This feature distinguishes our optimal IDNC solution from other IDNC solutions in the literature, where packet diversity has never been considered, to the best of our knowledge.

Second, the impact of feedback frequency on the performance of the IDNC scheme is investigated, the concept of semi-online feedback is introduced and optimal fully-online and  optimal semi-online IDNC schemes are devised.

Third, we derive lower and upper bounds on the throughput and decoding delay performance of IDNC schemes. Furthermore, we design the optimal IDNC coding algorithm, as well as its simplified alternatives that offer efficient performances with much lower computational complexities.

The performance of these algorithms is evaluated via extensive simulations under different settings of system parameters. The results illustrate the interactions among these parameters and can serve as simple implementation guidelines. Plenty of hands-on examples are also designed to demonstrate the proposed concepts, theorems, methodologies, and algorithms. In summary, this work can be a useful reference in the IDNC literature and can motivate further research.

\subsection{Additional Remarks}
IDNC can be divided into two categories, strict IDNC (S-IDNC) \cite{Rozner_Heuristic_clique,sadeghi:shams:traskov:2010} and general IDNC (G-IDNC) \cite{sameh:valaee:globecom:2010,le:dimakis:idnc_real}. Although they have the same system model and use similar dynamic algorithms, they differ in the sense that G-IDNC coded packets are allowed to include two or more new data packets for some receivers. However, this is not allowed in S-IDNC. In this paper, we focus on S-IDNC (or IDNC for short). The relationship and comparisons between S- and G-IDNC will be discussed when necessary.

S-IDNC problem is, to some extent, related to the index coding problem \cite{sprintson:min:2007,sprintson:algorithm:2008,sprintson:ic_nc_matroid:2010}, especially when memoryless decoding is considered in the index coding problem \cite{sprintson:min:2007,sprintson:algorithm:2008}.
Nevertheless, their problem formulations are different. A basic assumption in index coding is that a receiver who has successfully received a subset of packets but is still missing multiple packets can be considered as multiple receivers each wanting only one of the missing packets. However, such splitting is prohibited in S-IDNC, for it will violate the instantly decodable property of IDNC coded packets. Therefore, results of index coding and S-IDNC cannot be used interchangeably.

\section{System Model and Notations}
\subsection{Transmission Setup}
We consider a packet-based wireless broadcast scenario from one
sender to $N_T$ receivers. Receiver $n$ is denoted by $R_n$ and the
set of all receivers is  $\R_{N_T} = \{R_1, \cdots, R_{N_T}\}$. There are a
total of $K_T$ binary packets with identical length to be delivered to all receivers. Packet $k$
is denoted by $\p_k$ and the set of all packets is  $\P_{K_T} = \{\p_1,
\cdots, \p_{K_T}\}$. Sometimes we will refer to $\p_k$ as an original
data packet to distinguish it from a coded packet. Time is slotted
and in each time slot, one (coded or original data)  packet is
broadcast. The wireless channel between the sender and each receiver
is modeled as a memoryless erasure link with i.i.d. packet erasure probability of $P_e$. The results proposed in the paper can be
generalized, with proper modifications, to non-homogeneous
erasure links.

\subsection{Systematic Transmission Phase and Receivers Feedback}

Initially, the $K_T$ packets are transmitted uncoded once using $K_T$ time slots. This is the \emph{systematic transmission phase}. After this phase, each receiver provides feedback to the sender about the packets it has received or lost.\footnote{We assume that there exists an error-free feedback link between each receiver to the sender that can be used with appropriate frequency.} The number of packets that are not received by at least one receiver due to erasures is denoted by $K$ and their set is denoted by $\P_K = \{\p_1, \cdots, \p_{K}\}$, where $K\leqslant K_T$. The number of receivers that have not received all the $K_T$ packets  is denoted by $N$ and the set of these receivers is denoted by $\R_N=\{R_1,\cdots,R_N\}$, where $N\leqslant N_T$.

The complete state of receivers and packets can be captured by an $N \times K$ state feedback matrix (SFM)  $\mA$ (also known as receiver-packet incidence matrix~\cite{sadeghi:shams:traskov:2010}), where the element at row $n$ and column $k$ is denoted by $a_{n,k}$ and
\begin{equation}\label{eq:sfm}
  a_{n,k} =
   \begin{cases}
    1 & \text{if $R_n$  has lost $\p_k$}, \\
    0       & \text{otherwise.}
   \end{cases}
 \end{equation}
Based on the SFM, we define the notions of \emph{Wants} set \cite{sadeghi:adaptive_broadcast_2009,sameh:valaee:globecom:2010} for each receiver and \emph{Targeted} receivers for each packet:

\begin{Definition}
The Wants set of receiver $R_n$, denoted by $\W_n$, is the subset of packets in $\P_K$ which are lost at $R_n$ due to packet erasures. That is, $\W_n = \{\p_k: a_{n,k} = 1\}$.
\end{Definition}

\begin{Definition}
The Target set of a packet $\p_k$, denoted by $\T_k$, is the subset of receivers in $\R_N$ who want packet $\p_k$. That is, $\T_k = \{R_n: a_{n,k} = 1\}$. The size of $\T_k$ is denoted by $T_k$.
\end{Definition}

\begin{figure*}
\centering
\subfigure[State feedback matrix $mA$]{\includegraphics[width=0.23\linewidth]{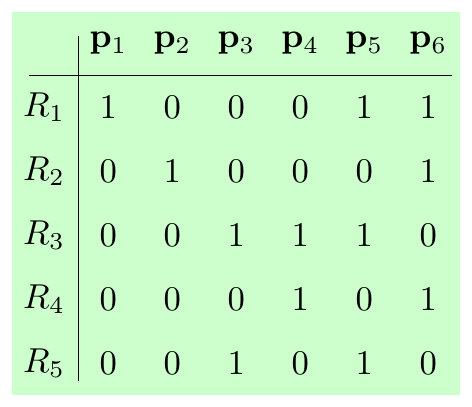}\label{fig:sfm_example}}\hspace{10pt}
\subfigure[Conflict matrix $\mC$]{\includegraphics[width=0.2\linewidth]{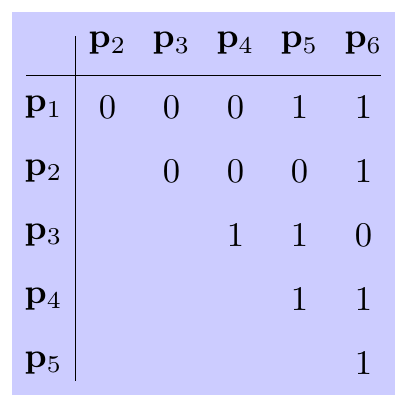}\label{fig:conflict_example}}\hspace{10pt}
\subfigure[Graph representation $\G$]{\includegraphics[width=0.21\linewidth]{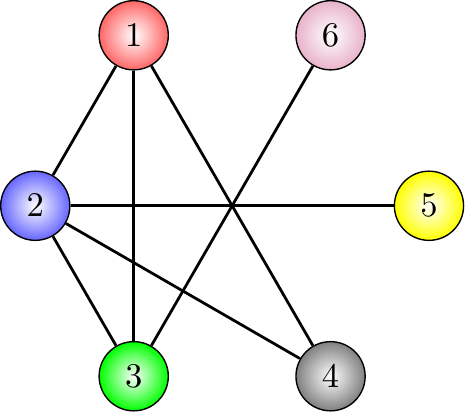}\label{fig:graph_example}}
\caption{An example of SFM and its matrix/graph representations. There are four maximal coding sets: $\{\p_1,\p_2,\p_3\}$, $\{\p_1,\p_2,\p_4\}$,$\{\p_2,\p_5\}$ and $\{\p_3,\p_6\}$. This SFM only has only one minimum collection: $\S=\{\{\p_1,\p_2,\p_4\},\{\p_2,\p_5\},\{\p_3,\p_6\}\}$.}
\label{fig:example}
\end{figure*}

\begin{Example}
Consider the SFM in \figref{fig:sfm_example}. There are $K=6$ packets and $N=5$ receivers after the systematic transmission phase. The Wants set of $R_1$ is $\W_1=\{\p_1,\p_5,\p_6\}$. The Target set of $\p_3$ is $\T_3=\{R_3,R_5\}$ and thus $T_3=2$.
\end{Example}

\subsection{Coded Transmission Phase}
In this subsection, we present some basic definitions and performance metrics related to IDNC. Then in the next subsection, we will briefly discuss existing models in the literature to deal with the IDNC problem.

After the systematic transmission phase and collecting receivers' feedback, the \emph{coded transmission phase} starts. In this phase, IDNC aims to satisfy the demands of all receivers by sending coded packets under two fundamental restrictions:
\begin{enumerate}
\item The sender uses the binary field $\F_2$ for linear coding;
\item Receivers do not store received coded packets for future decoding, i.e. memory is not required at the receivers;
\end{enumerate}

More precisely, the first restriction means that the $u$-th transmitted coded packet is of the form
\begin{equation}\label{eq:idnc:coding}
\X_u = \sum_{k = 1}^{K}\beta_{k,u} \p_k
\end{equation}
where $\beta_{k,u} \in \{0,1\}$ and the summation is bit-wise XOR $\oplus$. We denote by $\M_u$ the set of original data packets that have non-zero coefficients in $\X_u$, namely, $\M_u=\{\p_k:\beta_{k,u}=1\}$. $\M_u$ fully represents $\X_u$ and is called a coding set. Based on \eqref{eq:idnc:coding}, $\X_u$ can be one of the following for each receiver:
\begin{Definition}
A coded packet $\X_u$ is instantly decodable for receiver $R_n$ if
$\M_u$ contains only one original data packet from the Wants set $\W_n$ of $R_n$.
\end{Definition}
\begin{Definition}
A coded packet $\X_u$ is non-instantly decodable for receiver $R_n$ if $\M_u$ contains two or more original data packets from the Wants set $\W_n$ of $R_n$.
\end{Definition}
Due to restriction 2 above, a non-instantly decodable coded packet will be discarded by $R_n$ upon receiving.

\begin{Definition}
A coded packet $\X_u$ is non-innovative for receiver $R_n$ if $\M_u$ only contains original data packets not from the Wants set $\W_n$ of $R_n$. Otherwise, it is innovative.\end{Definition}
A non-innovative coded packet will be also discarded by $R_n$ upon receiving.

\begin{Example}
Consider the SFM in \figref{fig:sfm_example}. $\X=\p_4\oplus\p_5$ is instantly decodable for $R_1$ because the corresponding coding set $\M=\{\p_4,\p_5\}$ only contains one original data packet ($\p_5$) from the Wants set of $R_1$. Thus, $R_1$ can instantly decode $\p_5$ through the operation $\X\oplus\p_4$. $\X$ is also instantly decodable for $R_4$. However, $\X$ is non-instantly decodable for $R_3$ because both $\p_4$ and $\p_5$ are from the Wants set of $R_3$. $\X$ is non-innovative for $R_2$ because $R_2$ has both $\p_4$ and $\p_5$ already.
\end{Example}

The throughput and decoding delay performance of IDNC can be measured by \emph{the minimum number of coded transmissions} and \emph{the average packet decoding delay}, where:
\begin{Definition}
Given an SFM, the minimum number of coded transmissions, or equivalently the minimum block completion time, is the smallest possible number of IDNC coded transmissions required in order to satisfy the demands of all the receivers in the absence of any future packet erasures. This number is denoted by $\UIDNC$.
\end{Definition}

In the next section, we will further show that $\UIDNC$ cannot be reduced regardless of feedback frequency. Thus, we claim that $\UIDNC$ is the \emph{absolute} minimum number of coded transmissions. $\UIDNC$ indicates the best throughput performance, which can be calculated as $K_T/(K_T+\UIDNC)$. Such measure is important as it disentangles the effect of channel-induced packet erasures and algorithm-induced IDNC coded packet selections on the throughput of the system.

Next, we define the average packet decoding delay, $D$.

\begin{Definition}
Denote by $u_{n,k}$ the time slot in the coded transmission phase when original data packet $\p_k$ is decoded by receiver $R_n$, and let $u_{n,k}=0$ if $a_{n,k}=0$. Then:
\begin{equation}\label{eq:d_def}
D\triangleq\frac{1}{\sum_{k=1}^K T_k}\sum_{n=1}^N\sum_{k=1}^Ku_{n,k}
\end{equation}
\end{Definition}

\begin{Example}
Consider the SFM in \figref{fig:sfm_example}. Assume that four IDNC coded packets $\X_1=\p_1\oplus\p_2$, $\X_2=\p_3\oplus\p_6$, $\X_3=\p_4$, and $\X_4=\p_5$ are transmitted. Assuming erasure-free transmissions, all the receivers will be satisfied after four time slots. $\{u_{n,k}\}$ are summarized in Table \ref{tab:delay_sfm}. The block completion time is 4 and the average packet decoding delay is $(1\times2+2\times5+3\times2+4\times3)/12=2.5$. However, we have not discussed or determined yet if they are the best throughput and decoding delay performance of IDNC.

\begin{table}[H]
\centering
\begin{tabular}{c|cccccc}
~    &$\p_1$&$\p_2$&$\p_3$&$\p_4$&$\p_5$&$\p_6$\\\hline
$R_1$& 1    & 0    & 0    & 0    & 4    & 2\\\hline
$R_2$& 0    & 1    & 0    & 0    & 0    & 2\\\hline
$R_3$& 0    & 0    & 2    & 3    & 4    & 0\\\hline
$R_4$& 0    & 0    & 0    & 3    & 0    & 2\\\hline
$R_5$& 0    & 0    & 2    & 0    & 4    & 0\\\hline
\end{tabular}
\caption{The decoding delay of original data packets at the receivers}
\label{tab:delay_sfm}
\end{table}

\end{Example}

Optimizing the throughput and decoding delay performance in IDNC has been recognized as a highly non-trivial and computationally complex problem \cite{Rozner_Heuristic_clique,drinea:fragouli:keller:2009,sorour:valaee:2010, sameh:valaee:globecom:2010,sadeghi:shams:traskov:2010,sorour:valaee:2011}. Various approaches have been taken in the literature to solve this problem, which we now briefly discuss.

\subsection{S-IDNC versus G-IDNC}
One main model to capture IDNC constraints and determine coded packets is strict IDNC or S-IDNC \cite{Rozner_Heuristic_clique,drinea:fragouli:keller:2009,sadeghi:shams:traskov:2010}.  Imposing the S-IDNC constraint means that:

\begin{Definition}\label{def:S_IDNC_Constraint}
S-IDNC constraint: for every receiver $R_n$, the coding set $\M_u$ contains at most one original data packet from the Wants set of $R_n$. In other words, for every receiver, the coded packet $\X_u$ in \eqref{eq:idnc:coding} is either instantly decodable or non-innovative, but it is never non-instantly decodable.
\end{Definition}
In \cite{Rozner_Heuristic_clique}, it is shown that the S-IDNC constraint on the coded packets can be represented using an undirected graph with $K$ vertices corresponding to $K$ wanted original data packets. More details on the graphical representation of S-IDNC will be provided in \secref{sec:optimal_IDNC}.

In contrast, in the general IDNC or G-IDNC proposed in \cite{sorour:valaee:2010, sameh:valaee:globecom:2010} the S-IDNC constraint is relaxed by allowing the sender to send coded packets that are non-instantly decodable for a selected subset of receivers. If a receiver receives such a coded packet, it will discard that packet. In other words, in G-IDNC the sender is not restricted to send IDNC coded packets for all the receivers, but the receivers adhere to the IDNC decoding principle. Recently, a new type of G-IDNC  is proposed in \cite{neda:parastoo:o2idnc}, which further relaxes this constraint by allowing receivers to store non-instantly decodable packets for future decoding so that they are not wasted.

\begin{Example}
Consider the SFM in \figref{fig:sfm_example}. $\X=\p_2\oplus\p_5$ is a valid coded packet for both S-IDNC and G-IDNC. However, $\X=\p_4\oplus\p_5$ is a valid coded packet for G-IDNC but not for S-IDNC, since it is non-instantly decodable for $R_3$. In G-IDNC, when $R_3$ receives $\X=\p_4\oplus\p_5$, it will discard it.
\end{Example}

G-IDNC problem can also be modeled using an undirected graph where for each lost packet $\p_i$ of each receiver $R_j$ a vertex $v_{i,j}$ is added to the graph. The key operation of G-IDNC algorithms is to search for the largest \emph{maximal clique(s)}\footnote{In an undirected graph, all vertices in a clique are connected to each other with an edge. A clique is maximal if it is not a subset of a larger clique.} \cite{sorour:valaee:2010, sameh:valaee:globecom:2010,sorour:valaee:2011}. The size of G-IDNC graph is $O(NK)$, while the size of S-IDNC graph is $K$.

In the rest of this paper, our aim is to better understand and characterize the S-IDNC problem from both theoretical and implementation viewpoints. An important note is that the characteristics of S-IDNC cannot be directly extended to G-IDNC due to the fact that they construct and update their graphs in different ways, as will be explained in \secref{sec:optimal_IDNC}. Characterizing G-IDNC could be the objective of future research and is out of the scope of this paper. In the rest of the paper, when there is no ambiguity, we will simply refer to S-IDNC as IDNC.

\section{The Optimal IDNC}\label{sec:optimal_IDNC}

IDNC constraints of an SFM $\mA$ can be represented by an undirected graph $\G(\V,\E)$ with $K$ vertices. Each vertex $v_i\in\V$ represents a wanted original data packet $\p_i$. Two vertices $v_i$ and $v_j$ are connected by an edge $e_{i,j}\in\E$ if $\p_i$ and $\p_j$ are not jointly wanted by any receiver \cite{Rozner_Heuristic_clique}. This graph model, however, has only been employed in the literature to heuristically find IDNC coding solutions \cite{Rozner_Heuristic_clique}. In this section, we will first revisit this graph model by constructing its equivalent matrix and set models. Then, we will use these models to rigorously prove some theorems about the minimum block completion time, $\UIDNC$. The key difference between S-IDNC and G-IDNC will become clear after the proofs. Based on these theorems, we will discuss the effect of feedback frequency on IDNC throughput and propose the optimal IDNC schemes with fully- or semi-online feedback. Although some similar concepts and results exist in the graph theory literature \cite{Graph_coloring,Graph_theory}, their compilation, presentation and more importantly interpretation in the IDNC context is new, to the best of our knowledge. We will highlight the similarities, differences, and new results as appropriate.
\vspace{-10pt}

\subsection{IDNC Modeling}
In this subsection, we construct the matrix and set models of IDNC and demonstrate their relationship with its graph model. The construction is based on the concepts of conflicting and non-conflicting original data packets, defined as follows.

\begin{Definition}\label{def:conf_packet}
Two original data packets $\p_i$ and $\p_j$ conflict with each other if both belong to the Wants set, $\W_n$, of at least one receiver such as $R_n$. Mathematically, we can denote a conflict between $\p_i$ and $\p_j$ by $\p_i\conf  \p_j$, where $\p_i \conf \p_j \eqv \exists n: \{\p_i,\p_j\}\subseteq\W_n$. $\p_i$ and $\p_j$ do not conflict otherwise.
\end{Definition}

It is clear that to avoid non-instantly decodable coded packets, two conflicting original data packets $\p_i$ and $\p_j$ cannot be coded together. The equivalent of such conflict in the graph model is the absence of an edge between $v_i$ and $v_j$ \cite{Rozner_Heuristic_clique,Graph_theory}. On the other hand, two non-conflicting data packets $\p_i$ and $\p_j$ have their respective vertices $v_i$ and $v_j$ connected.

The conflict states of all the original data packets can be fully described by a triangular conflict matrix $\mC$ of size $K(K-1)/2$:

\begin{Definition}
A fully-square conflict matrix of size $K^2$ is a binary-valued matrix with element at row $i$ and column $j$ denoted by $c_{i,j}$ corresponding to the conflict state of packets $\p_i$ and $\p_j$. In particular, $c_{i,j} = 1$ if  $\p_i\conf  \p_j$ and  $c_{i,j} = 0$ otherwise.
Due to the symmetry of conflict between packets and noting that $c_{i,i} = 0$, $\forall \p_i$, we can reduce the fully-square matrix to a triangular matrix $\mC$ of size $K(K-1)/2$. From now on, by conflict matrix, we mean the reduced triangular matrix $\mC$.
\end{Definition}

The conflict matrix and the graph model of the SFM in \figref{fig:sfm_example} are presented in \figref{fig:conflict_example} and \figref{fig:graph_example}, respectively. Note that in dealing with the conflict matrix, we are not concerned with the receivers who may need a certain packet. In fact, it is not difficult to show that two or more SFMs can have the same conflict matrix. Unless otherwise stated, it suffices to deal with conflict matrix $\mC$ for design and analysis of IDNC, instead of SFM $\mA$.

We now define the key concept of \emph{maximal coding set} that is allowed for IDNC transmission.

\begin{Definition}
A maximal coding set, $\M$, is a set of original data packets which simultaneously hold the following two properties: 1) their XOR coded packet $\X$ satisfies IDNC constraint in Definition~\ref{def:S_IDNC_Constraint}, i.e. $\X$ is either instantly decodable or non-innovative for every receiver $R_n\in\R_N$; 2) addition of any other original data packet from $\P_K\setminus \M$ to $\M$ will make the resulted $\X$ non-instantly decodable for at least one receiver in $\R_N$.
\end{Definition}
\begin{Example}\label{exmp:all_sets}
Consider a coding set $\M=\{\p_2,\p_5\}$ of the SFM in \figref{fig:sfm_example}. Its corresponding coded packet is $\X=\p_2\oplus\p_5$. $\X$ is instantly decodable for $R_{\{1,2,3,5\}}$ and is non-innovative for $R_4$. One can verify that adding any other original data packet from $\P_K\setminus\M=\{\p_1,\p_3,\p_4,\p_6\}$ to $\M$ will make $\X$ non-instantly decodable for at least one receiver. Hence, $\M$ is a maximal coding set. Through exhaustive search, one can find all remaining maximal coding sets: $\{\p_1,\p_2,\p_3\}$, $\{\p_1,\p_2,\p_4\}$, $\{\p_3,\p_6\}$.
\end{Example}
The equivalent of maximal coding sets in the IDNC graph model is known as maximal cliques \cite{Graph_theory}. Therefore, we use $\M$ to denote both a maximal coding set and a maximal clique.

For reasons that become clear at the end of this subsection, we ensure that in each IDNC coded transmission, the sender will code \emph{all and not a subset of} the original data packets in a maximal coding set. To satisfy the demands of all the receivers, the sender has to transmit coded packets from an appropriately chosen collection of maximal coding sets. To achieve this, each original data packet should appear at least once in this collection. This condition can be formally represented as the \emph{diversity constraint}, where diversity of a packet is defined as:
\begin{Definition}
The diversity of an original data packet $\p_i$ within a collection of maximal coding sets is denoted by $d_i$ and is the number of maximal coding sets in which it appears.
\end{Definition}
\begin{Definition}
A collection of maximal coding sets satisfies the diversity constraint iff every original data packet has a diversity of at least one within this collection.
\end{Definition}

Given all the maximal coding sets of a conflict matrix $\mC$, there exists at least one collection which satisfies the diversity constraint (in the extreme case all the maximal coding sets include all the original data packets). The size of the collection is the number of maximal coding sets in it. We then define the \emph{minimum collection} and its size as follows:
\begin{Definition}
A collection of maximal coding sets is minimum if there does not exist any other collection which satisfies the diversity constraint with a smaller size. The size of the minimum collection is called the minimum collection size.
\end{Definition}

This number, as we will prove in the next subsection, is exactly the minimum number of coded transmissions, $\UIDNC$. We thus denote a minimum collection by $$\S=\{\M_1,\cdots,\M_{\UIDNC}\}$$
If the $\UIDNC$ maximal coding sets in $\S$ are sent using $\UIDNC$ time slots, in the absence of packet erasures, the demands of all receivers will be satisfied. Here by ``sending a maximal coding set'', we mean that the corresponding coded packet is generated and sent.

A problem in the graph theory which is somewhat \emph{similar} to finding a minimum collection of maximal coding sets is the minimum clique cover problem~\cite{Rozner_Heuristic_clique,Graph_theory,Graph_coloring}. In this problem, a graph $\G$ is partitioned into disjoint cliques and the partitioning solution that results into the smallest number of disjoint cliques is referred to as minimum clique cover solution of the problem. Furthermore, it is worth noting that the cardinality of the minimum clique cover solution is equal to the chromatic number of the complementary graph \footnote{The complementary graph $\cG$ has opposite vertex connectivity to $\G$.} of $\G$, denoted by $\chi(\cG)$ \cite{Graph_theory}. However, there is a difference between the minimum clique cover problem and our minimum collection finding problem, as cliques do not overlap in the minimum clique cover problem. That is, the cliques are not necessarily maximal and each vertex appears in only one clique. This would be equivalent to choosing a minimum collection of coding sets in our IDNC model where all original data packets have a diversity equal to one. This would not change $\UIDNC$. However, it can have a serious adverse impact on IDNC's robustness to erasures, which in turn degrades the IDNC overall throughput and decoding delay performance. Consequently, it is desirable to choose a minimum collection of \emph{maximal} coding sets that, while satisfying $\UIDNC$, provides as many \emph{packet diversities} as possible. The minimum collection and the importance of packet diversity is illustrated with the following example.

\begin{Example}\label{exmp:min_collection}
Having all the maximal coding sets of SFM in \figref{fig:example} obtained in Example \ref{exmp:all_sets}, one can easily verify that the only minimum collection is $$\S=\{\{\p_1,\p_2,\p_4\},\{\p_2,\p_5\},\{\p_3,\p_6\}\}$$ By sending these three coding sets in $\S$ using $\UIDNC=3$ transmissions, the demands of all the receivers will be satisfied in the absence of packet erasures. All the original data packets in $\S$ have a diversity of one, except $\p_2$ which has a diversity of 2, i.e. $d_2=2$. Now, let us assume that there is a packet erasure probability of $P_e=0.2$ in the transmission links between the sender and the receivers. Under this scenario, with these  three coded transmissions, the probability of  $\p_2$ being lost at its targeted receiver $R_2$ (due to erasures) will be $P_e^2=0.04$. This probability is much lower then that of other original data packets, which will be equal to  $P_e=0.2$.
\end{Example}
\begin{Remark}
The problem of finding all the maximal cliques and the problem of minimum clique cover for an undirected graph are both NP-complete \cite{Graph_theory,Graph_coloring,Karp_NP_problems}. This is also true for S-IDNC because the S-IDNC graph does not have any special structural properties. A similar statement can be found in \cite{le:dimakis:idnc_real} for G-IDNC. Since a minimum collection of a S-IDNC conflict matrix can be reduced to a minimum clique cover solution of a S-IDNC graph by reducing the diversity of all data packets to one, the problem of finding minimum collections in S-IDNC is at least NP-complete. Its exact and simplified algorithms will be presented in \secref{sec:implementations}.
\end{Remark}

\subsection{The Equivalence of $\UIDNC$, the Minimum Collection Size, and $\chi(\cG)$}

In this subsection, we prove that the three numbers: 1) the minimum number of coded transmissions ($\UIDNC$), 2) the minimum collection size of the conflict matrix, and 3) the chromatic number of the complementary graph ($\chi(\cG)$), are identical. Based on this we propose two important remarks.

For a given conflict matrix $\mC$, the equivalence between its $\UIDNC$ and minimum collection size can be proved by induction using the following theorem:

\begin{Theorem}\label{theo:UIDNC}
Upon successful reception of a maximal coding set $\M$ of $\mC$ by all its targeted receivers, the minimum collection size of the updated $C$, denoted by $\mC'$, is at least $\UIDNC-1$.
\end{Theorem}

This theorem holds if the following two theorems hold:
\begin{Theorem}\label{theo:chromatic}
The minimum collection size of a conflict matrix $\mC$ with a graph model $\G$ equals the chromatic number of the complementary graph of $\G$, $\chi(\cG)$.
\end{Theorem}
\begin{Theorem}\label{theo:chromatic_minus}
Suppose $\M$ is a maximal clique in $\G$ and the chromatic number of $\overline{\G}$ is $\UIDNC$. By removing $\M$ from $\G$ we obtain an updated graph $\G'$. The chromatic number of $\cGn$ is at least $\UIDNC-1$. More precisely, if $\M$ belongs to a minimum collection of $\mC$, $\chi(\cGn)=\UIDNC-1$, while if $\M$ does not belong to any minimum collection of $\mC$, $\chi(\cGn)=\UIDNC$.
\end{Theorem}
Since the $\G'$ in Theorem \ref{theo:chromatic_minus} is indeed the graph model of $\mC'$ in Theorem \ref{theo:UIDNC}, we conclude that Theorem \ref{theo:UIDNC} holds if Theorem \ref{theo:chromatic} and \ref{theo:chromatic_minus} hold. The proofs of Theorem \ref{theo:chromatic} and \ref{theo:chromatic_minus} are provided in \ref{app:1} and \ref{app:2}, respectively.

\begin{Remark}
The above theorems apply to S-IDNC, but not G-IDNC. The reason is that unlike S-IDNC graph,  G-IDNC graph is not \textbf{static}. That is, by removing a clique from the G-IDNC graph, new edges may be added to the remaining vertices, which breaks Theorem \ref{theo:chromatic_minus}. In contrast, removing any clique from the S-IDNC graph will never change the connectivity of the remaining vertices. In other words, G-IDNC problem does not have its dual \textbf{static} minimum clique cover problem.
\end{Remark}

\begin{Remark}\label{remark:suboptimal}
Heuristic algorithms are suboptimal because they cannot guarantee $\UIDNC$. They may choose a maximal coding set $\M$ which is, though large, not included in any minimum collection of $\mC$. Then Theorem \ref{theo:chromatic_minus} indicates that, even if $\M$ is successfully received by all its targeted receivers, the chromatic number of the updated $\overline{\G'}$ is still $\UIDNC$ and thus $\UIDNC$ more transmissions are needed. Below is an example.
\end{Remark}

\begin{Example}
Consider the maximal coding sets in Example \ref{exmp:all_sets}. A suboptimal IDNC algorithm might choose $\{\p_1,\p_2,\p_3\}$, which does not belong to the only minimum collection $\S=\left\{\{\p_1,\p_2,\p_4\},\{\p_2,\p_5\},\{\p_3,\p_6\}\right\}$. Even if this set is successfully received by all its targeted receivers, still three more transmissions are needed to be able to deliver $\p_4,\p_5$ and $\p_6$ to the receivers. In total, there will be at least four transmissions, which is greater than $\UIDNC=3$.
\end{Example}

This example motivates the concept of optimal IDNC schemes, which will be presented next.

\subsection{The Optimal Fully- and Semi-online IDNC Schemes}\label{sec:optimal_schemes}

\subsubsection{The optimal fully-online IDNC scheme}
In a fully-online IDNC scheme, the sender collects feedback from all receivers in every time slot to update the SFM and the corresponding conflict matrix. A coding set is then chosen and its coded packet is generated and broadcast. To minimize the number of coded transmissions and to reduce the decoding delay, this coding set must satisfy the following three conditions:
\begin{enumerate}
\item It should be a maximal coding set;
\item It should be chosen from a minimum collection $\S$ of the updated conflict matrix; and
\item It should target the largest number of receivers among all the maximal coding sets in $\S$.
\end{enumerate}

We define such a fully-online IDNC scheme as the \emph{optimal fully-online IDNC scheme} in terms of throughput.

\subsubsection{The optimal semi-online IDNC scheme}
\begin{figure}
\centering
\includegraphics[width=\linewidth]{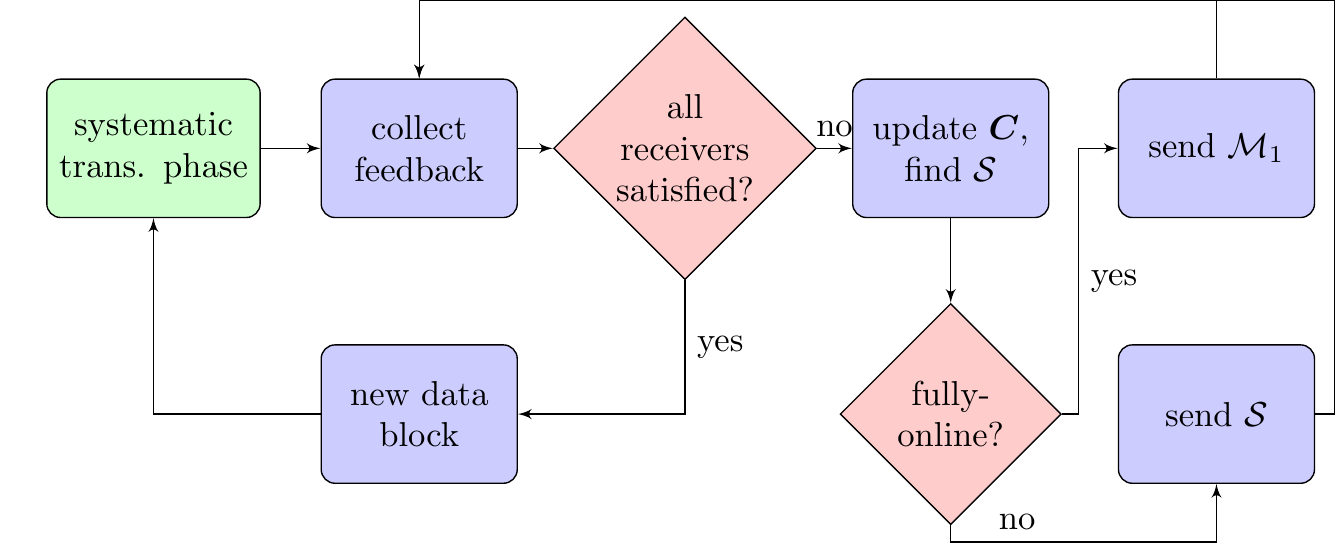}
\caption{The optimal fully- and semi-online IDNC schemes.}
\label{fig:full_semi_flow}
\end{figure}

According to Theorem \ref{theo:UIDNC}-\ref{theo:chromatic_minus}, collecting fully-online feedback during $\UIDNC$ transmissions cannot reduce the total number of coded transmissions to below $\UIDNC$, even in the best case scenario of erasure-free packet reception. Hence, as a variation of existing IDNC schemes in the literature, we propose to reduce feedback frequency to \emph{semi-online}, where the SFM $\mA$ is updated in rounds. For example, feedback is collected after $\UIDNC$ coded packets from a selected minimum collection $\S$ have been transmitted and so on. We define this scheme as the optimal IDNC scheme in terms of throughput when feedback frequency is semi-online, or simply \emph{the optimal semi-online IDNC scheme}. We refer to the minimum collection $\S$ as the \emph{optimal semi-online IDNC solution}. Maximal coding sets in $\S$ are properly ordered so that those targeting more receivers are assigned with smaller subscripts and sent first.  \figref{fig:full_semi_flow} illustrates the process of the proposed optimal IDNC schemes.

We then define the minimum average packet decoding delay of the proposed IDNC schemes:
\begin{Definition}
We denote by $\DIDNC$ the minimum average packet decoding delay of the proposed fully- and semi-online IDNC scheme. It is achieved if the maximal coding sets in the optimal semi-online IDNC solution $\S$ are broadcast in the absence of packet erasures, and is calculated as:
\begin{equation}\label{eq:didnc_def}
\DIDNC\triangleq\frac{1}{\sum_{k=1}^K T_k}\sum_{k=1}^Ku_kT_k
\end{equation}
where $u_k$ is the index of the first maximal coding set in $\S$ which contains $\p_k$.
\end{Definition}
Compared with \eqref{eq:d_def}, the decoding delays of an original data packet $\p_k$ at its targeted receivers, i.e., $\{u_{n,k}\}$, are unified to $u_k$ here because all these receivers can decode $\p_k$ in the same time slot if there is no packet erasure. It is noticed that by ``minimum'' we mean the smallest possible  average packet decoding delay of the proposed (throughput) optimal IDNC schemes in the absence of packet erasures. $\DIDNC$ is not necessarily the optimal average packet decoding delay that IDNC can offer, finding which is still an open problem. Indeed, as we shall see in \secref{sec:simulations}, a suboptimal IDNC scheme in terms of throughput may achieve a better decoding delay.

It is also noticed that, since the initial SFM $\mA$ is a $K_T\times K_T$ all-one matrix, the systematic transmission phase is a special semi-online IDNC round, which requires the $K_T$ original data packets to be sent uncoded using $\UIDNC=K_T$ transmissions.

\subsubsection{Comparisons}
In addition to making the throughput and delay analysis of IDNC tractable, a lower feedback frequency can be advantageous in practical implementations of IDNC where the use of reverse link is costly and involves transmission of some control overheads. Another practical attraction is that it also avoids solving the IDNC coding problem in every time slot. However, this comes at the potential cost of degradation in the overall system throughput in semi-online IDNC, as we explain next.

Imagine an IDNC scheme in the presence of erasures. In the fully-online feedback case, $\mA$ is updated before every coded transmission, so the coded packet is chosen from the minimum collection of the actual $\mA$ at the receivers. However, in the semi-online feedback case, the sender does not update $\mA$ until the round for $\S=\{\M_1,\M_2,\cdots,\M_{\UIDNC}\}$ is complete. Here $\M_2,\cdots,\M_{\UIDNC}$ belong to the minimum collection of the $\mA$ last revealed to the sender, but not necessarily belong to the actual $\mA$ at the receivers. If this is the case, these coded packets can become throughput inefficient. Intuitively, we expect the gap between semi- and fully-online schemes to be small when packet erasure probability is low (in the extreme case where the packet erasure probability is zero, the two schemes perform the same). In any case, the throughput and delay analysis of semi-online IDNC scheme serves as a worst-case scenario for an optimal fully-online IDNC with packet erasures.

\section{Throughput Bounds}\label{sec:throughput_bounds}
The findings in the last section are important because they enable theoretical analysis on the achievable throughput and decoding delay of IDNC. For throughput, $\UIDNC$ is equal to the chromatic number of the complementary IDNC graph. For decoding delay, $\DIDNC$ is the average decoding delay of the proposed optimal semi-online IDNC solution. We note, however, that there is no explicit formula to calculate the optimal $\UIDNC$. It can only be found via algorithmic implementations that can be computationally complex, as will be discussed in Section \ref{sec:implementations}. Therefore, it is desirable to have some bounds on $\UIDNC$ that can be more easily calculated or algorithmically found. This is the aim of this section. It is particularly useful and can find its application in, e.g., adaptive network coding systems which choose among IDNC and other network coding techniques to meet the throughput and decoding delay requirements. Since the calculation of $\DIDNC$ depends on $\UIDNC$ as indicated by \eqref{eq:didnc_def}, we will first derive bounds on $\UIDNC$ in this section and then on $\DIDNC$ in the next section. We start with the review of existing results in graph theory and then propose useful bounds in IDNC context.

\subsection{Results in Graph Theory}\label{sec:u_graph_result}
Given a set of system parameters $\{K_T,N_T,P_e$\}, the complementary IDNC graph $\cG$ after the systematic transmission phase can be modeled as the classic Erdos-Renyi random graph \cite{Random_graph}. In this model, there are $K_T$ vertices and any two of them are connected by an undirected edge with i.i.d. probability of $P_c$. In the context of IDNC, $P_c$ is the probability that two original data packets conflict with each other and can be calculated as:
\begin{equation}
P_c=1-(1-P_e^2)^{N_T}
\end{equation}

The chromatic number of this random graph model has the following property \cite{bollobas:chromatic,sorour:valaee:2011}:
\begin{Lemma}
Almost every random graph with $K_T$ vertices and vertex connection probability of $P_c$ has chromatic number of:
\begin{equation}\label{eq:chromatic_number}
\left(\frac{1}{2}+o(1)\right)\log\left(\frac{1}{1-P_c}\right)\frac{K_T}{\log K_T}
\end{equation}
where $o(1)$ approaches zero with increasing $K_T$.
\end{Lemma}

Since $\UIDNC=\chi(\cG)$, this lemma could be used to calculate the mean of $\UIDNC$ under any set of $\{K_T,N_T,P_e\}$. However, it is only asymptotically accurate for large $K_T$. Since in IDNC systems $K_T$ may not be very large, \eqref{eq:chromatic_number} does not provide sufficient accuracy.

In graph theory, $\chi(\cG)$ is bounded as:
\begin{equation}\label{eq:u_graph_bounds}
\w(\cG)\leqslant\chi(\cG)\leqslant\Delta(\cG)+1
\end{equation}
where $\w(\cG)$ is called the clique number of $\cG$ and is the size of the maximum (the largest maximal) clique in $\cG$, and $\Delta(\cG)$ is the largest vertex degree of $\cG$, i.e., the largest number of edges incident to any vertex in $\cG$. As we will show later, while $\w(\cG)$ is a tight lower bound, the upper bound $\Delta(\cG)+1$ is very loose and is not useful for IDNC framework. In the next two subsections, we will derive useful loose/tight lower/upper bounds on $\UIDNC$, respectively. The loose bounds are easy to calculate and they reveal the limits of IDNC, while the tight bounds are more computationally involved, but nevertheless are shown numerically to be accurate estimates of $\UIDNC$. 
\vspace{-1em}
\subsection{Loose Bounds}
In this subsection, we find the smallest and largest possible $\UIDNC$ of all the conflicts matrices which have a size of $K$ and $M_0$ zero entries, with their set denoted by $\C(K,M_0)$. The results are our loose lower and upper bounds and are denoted by $U(K,M_0)^-$ and $U(K,M_0)^+$, respectively. They reveal the throughput limits of IDNC for any given $K$ and $M_0$ and are important references for practical/heuristic IDNC coding algorithm design: any algorithm offering $\UIDNC$ above the upper bound or below the lower bound is throughput inefficient or non-instantly decodable, respectively.
\setcounter{equation}{8}
\begin{figure*}[t]
\begin{subequations}\label{eq:Ct_min_demo}
\begin{align}
\S&=\{\{\p_1\},\{\p_2\},\{\p_3\},\{\p_4\},\{\p_5\}\}~~~~~\UIDNC=5\label{eq:Ct_min_0}\\\label{eq:Ct_min_1}
\S&=\{\{\p_1,\p_2\},\{\p_3\},\{\p_4\},\{\p_5\}\}~~~~~1\times1=1 ~~\textrm{zero needed},~\UIDNC=4\\\label{eq:Ct_min_2}
\S&=\{\{\p_1,\p_2\},\{\p_3,\p_4\},\{\p_5\}\}~~~~~1\times1=1 ~~\textrm{zero needed},~\UIDNC=3\\\label{eq:Ct_min_3}
\S&=\{\{\p_1,\p_2,\p_5\},\{\p_3,\p_4\}\}~~~~~2\times1=2 ~~\textrm{zeros needed},~\UIDNC=2\\\label{eq:Ct_min_4}
\S&=\{\{\p_1,\p_2,\p_3,\p_4,\p_5\}\}~~~~~2\times3=6 ~~\textrm{zeros needed},~\UIDNC=1
\end{align}
\end{subequations}
\end{figure*}
\subsubsection{$U(K,M_0)^+$}
The general intuition here  is \emph{trying our best to waste the coding opportunities brought by the $M_0$ zeros}. We first note that for any given original data packet, there are $K-1$ entries in $\mC$ about the conflict of that packet with all other packets. When $M_0=0$, there is no coding opportunities, so $U(K,M_0=0)^+=K$. When $M_0\in[1,K-1]$, we can assign all zeros to the entries about the same original data packet, say $\p_1$. But $\UIDNC$ remains $K-1$ because $\p_2,\cdots,\p_K$ have to be transmitted separately. After $K-1$ zeros have been exhausted, there are $K-2$ entries in $\mC$ about every packet other than $\p_1$. Thus when $M_0\in[K,(K-1)+(K-2)]$, we can assign these extra $K-2$ zeros to the entries about the same original data packet, say $\p_2$, and $\UIDNC$ remains $K-2$ because $\p_3,\cdots,\p_K$ have to be transmitted separately. This iterative process indicates that $U(K,M_0)^+$ decreases in a staircase way with $M_0$. The relationship can be written as:
\setcounter{equation}{7}
\begin{equation}\label{eq:Ct_max_eq}
U(K,M_0)^+=\begin{cases}
                K,& M_0=0\\
                K-1,& M_0\in[1,K-1]\\
                K-2,& M_0\in[K,2K-3]\\
                \vdots&\vdots\\
                1,   & M_0=K(K-1)/2
\end{cases}
\end{equation}
\setcounter{equation}{9}
One can easily verify that the proposed loose upper bound is much tighter than $\Delta(\cG)+1$ because the largest possible $\Delta(\cG)+1$ is always $K$ when $M_0\in[0,K(K-1)/2-(K-1)]$.

\subsubsection{$U(K,M_0)^-$}
The intuition here is \emph{making the best of the coding opportunities brought by the $M_0$ zeros}. In other words, we should \emph{use as few zeros as possible to reduce $\UIDNC$ by one}. When $M_0=0$, no original data packets can be coded together. Thus $\S=\{\{P_1\},\{P_2\},\cdots,\{P_K\}\}$ and $\UIDNC=K$. We then reduce $\UIDNC$ iteratively. In each iteration, $\UIDNC$ can be reduced by 1, i.e., the size of $\S$ can be reduced by 1, if we can merge two maximal coding sets in $\S$ together. Since any two packets from different maximal coding sets conflict, to merge two maximal coding sets of size, say $m$ and $n$, together, we need $mn$ zeros. In order to use as few zeros as possible, we always pick two smallest sets which minimize $mn$. Hence, in each iteration, it is impossible to reduce $\UIDNC$ by 1 until $mn$ new zeros are added to $\mC$.  This iterative process provides the lower-bound $U(K,M_0)^-$. Similar to the upper-bound, the lower bound also decreases in a staircase way with $M_0$. Below is an example with $K=5$:
\begin{Example}\label{exmp:UIDNC_min}

When we have an all-one conflict matrix ($M_0=0$), no packets can be coded together, and thus $\UIDNC=5$, as in (\ref{eq:Ct_min_0}). Then in the first iteration, the size of $\S$ can be reduced by 1 by merging $\{\p_1\}$ and $\{\p_2\}$ together, which requires one zero, as in (\ref{eq:Ct_min_1}). In the second iteration, the size of $\S$ can be reduced by one by merging $\{\p_3\}$ and $\{\p_4\}$ together, which requires 1 zero, as in (\ref{eq:Ct_min_2}). In the third iteration, $2\times 1=2$ zeros are needed to merge the two smallest maximal coding sets $\{\p_1,\p_2\}$ and $\{\p_5\}$ together because we have to resolve the conflicts between $\p_1$ and $\p_5$ and between $\p_2$ and $\p_5$. In the last iteration, $3\times2=6$ zeros are needed to merge $\{\p_1,\p_2,\p_5\}$ and $\{\p_3,\p_4\}$ together. After that, all the 10 entries in $\mC$ become zeros and $\UIDNC$ becomes 1. $U(K,M_0)^-$ in this example can thus be expressed as:
\vspace{-5pt}
\begin{equation}
\UIDNC(5,M_0)^-=\begin{cases}
5,&M_0=0\\
4,&M_0=1\\
3,&M_0\in[2,3]\\
2,&M_0\in[4,9]\\
1,&M_0=10
\end{cases}
\end{equation}
\end{Example}

This relationship can be approximated by a single formula which is derived by Geller in \cite{Geller_lower_bound} using a different approach:
\begin{equation}\label{eq:Geller}
U(K,M_0)^-\approx\left\lceil\frac{K^2}{K+2M_0}\right\rceil,
\end{equation}
where $\lceil m\rceil$ denotes the smallest integer greater than $m$. One can easily verify that the proposed $U(K,M_0)^-$ is slightly tighter than the bound in \eqref{eq:Geller}.

\subsection{Tight bounds}\label{sec:u_tight}
Because $\chi(\cG)$ is well lower bounded by $\w(\cG)$ and $\UIDNC=\chi(\cG)$, our tight lower bound on $\UIDNC$ is defined as $U^-\triangleq\w(\cG)$. It can be identified by finding the maximum clique in $\cG$. We then find a tight upper bound, denoted by $U^+$.

\subsubsection{$U^+$}\label{sec:throughput_upper}
Our tight upper bound on $\UIDNC$ is derived using an iterative operation on a graph $\G$, denoted by $\f(\G)$. It iteratively outputs the maximum clique in $\G$ and then deletes it from $\G$ until $\G$ becomes empty. Mathematically:
\begin{align}\label{eq:upper_operation}
&\{\M'_0,\M'_1,\cdots\}=\f(\G),\\\nonumber
&\textrm{where~~} \M'_i:\M'_i\in\G_{i},|\M'_i|=\w(\G_{i}),\\\nonumber
&\textrm{and~~} \G_0=\G,~\G_i=\G_{i-1}\setminus \M'_{i-1} \mathrm{~for~}i>0
\end{align}

The resulted cliques actually form a partition of $\G$. Hence, if the initial $\G$ is an instance of IDNC graph, the resulted cliques form a semi-online IDNC solution, denoted by $S_{U^+}$ with cardinality $U^+$. The minimum block completion time ($\UIDNC$) of this IDNC instance is thus upper bounded by $U^+$.
\begin{Remark}
The derivations of both $U^-$ and $U^+$ rely on finding the maximum clique in an IDNC graph, which is NP-complete \cite{Graph_theory,Graph_coloring,Karp_NP_problems}. In \secref{sec:implementations}, we will propose a heuristic clique-finding algorithm, which will in turn enable finding some heuristic bounds on $\UIDNC$. These heuristic bounds are still provable, but they are suboptimal because the heuristically found maximal cliques are not necessarily maximum.
\end{Remark}

\begin{figure*}[ht]
\centering
\includegraphics[width=0.7\linewidth]{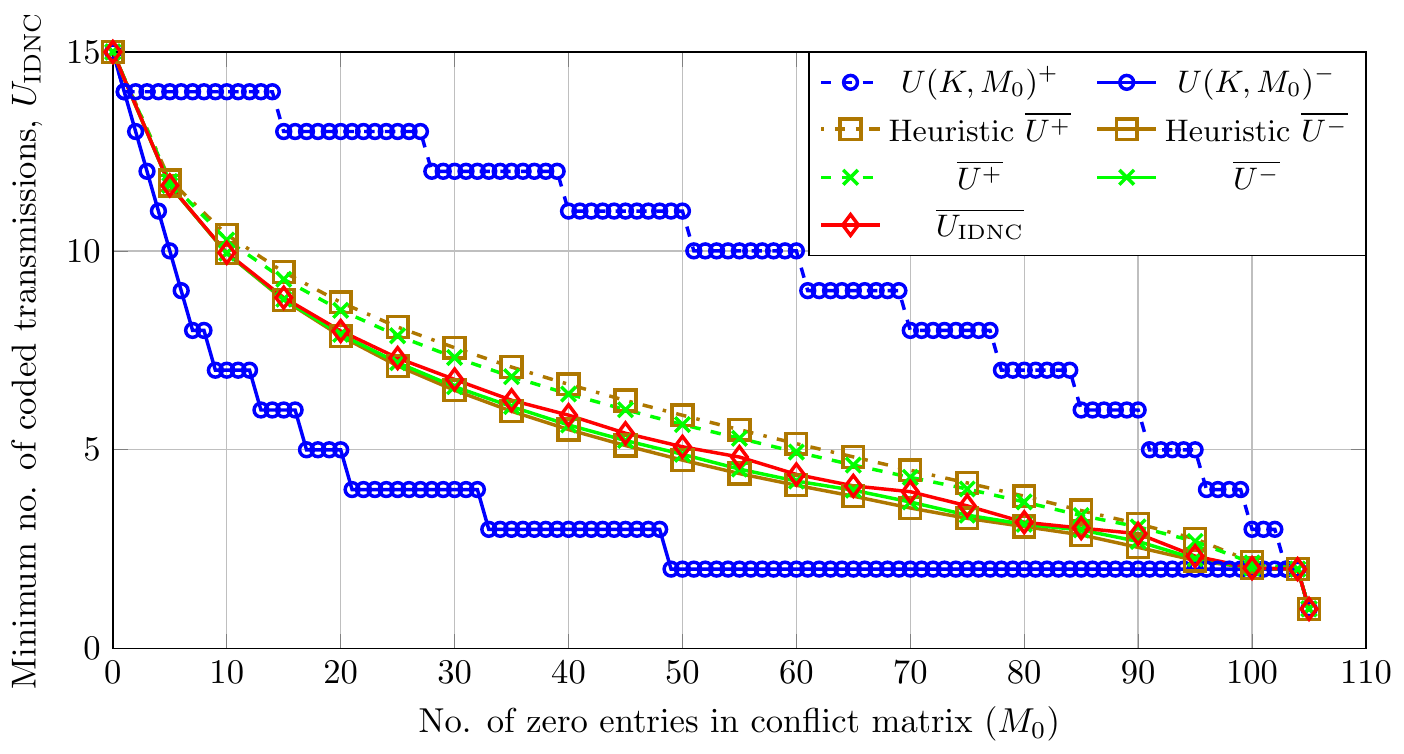}
\caption{Upper and lower bounds on the throughput with $K=15$ wanted original data packets.}\label{fig:u_bounds_15}
\end{figure*}
\subsection{The Average Bounds on $\UIDNC$ and Their Tightness}\label{sec:u_average}

Based on the bounds we have derived for any instance of conflict matrix, we can calculate the average bounds over all the conflict matrices in $\C(K,M_0)$ so that the average bounds can be explicitly expressed as a function of $\{K,M_0\}$.

Since the loose bounds are already functions of $\{K,M_0\}$, the focus is on the average tight bounds, i.e., $\overline{U^-}$ and $\overline{U^+}$. They can be obtained by listing all the conflict matrices in $\C(K,M_0)$, calculating their bounds, and then making the average. However, this is usually unrealistic, since there are $\binom{(K^2-K)/2}{M_0}$ possible conflict matrices, which are prohibitively large even when $K$ and $M_0$ are not so large. Hence, averaging over ``all'' conflict matrices is replaced by Monte Carlo averaging, where instances of conflict matrix are generated by assigning random permutations of $M_0$ zeros and $(K^2-K)/2-M_0$ ones to the conflict matrix.

We present the average bounds under $K=15$ original data packets and $M_0\in[0,105]$ in \figref{fig:u_bounds_15}. The optimal $\UIDNC$, obtained using the method in \secref{sec:optimal_algorithms}, is also averaged and plotted as a reference. It is denoted by $\overline{\UIDNC}$. It decreases gradually with the increasing $M_0$, and so do the tight bounds $\overline{U^-}$ and $\overline{U^+}$. The gap between the tight bounds and the optimal one is marginal, with a value of less than 0.5 transmission on average for all $M_0$. They are much tighter than the loose ones, which decrease in a stair-case way with increasing $M_0$.

\def\umin{U_{\min}}
\def\umax{U_{\max}}
\def\dmin{D_{\min}}
\def\dmax{D_{\max}}
\def\tt{T_{\textrm{total}}}
\def\M{\mathcal M}

\section{Decoding Delay Bounds}\label{sec:delay}

According to its definition in \eqref{eq:didnc_def}, the minimum average packet decoding delay $\DIDNC$ of an SFM $\mA$ is decided by the optimal semi-online IDNC solution $\S$ of the corresponding conflict matrix $\mC$ and the number of targeted receivers of all the original data packets $\{T_k\}$. Deriving lower/upper bounds on $\DIDNC$ of an SFM is thus equivalent to finding two instances of $\S$ which offer the best/worst possible decoding delays, respectively. We first discuss what instances will yield such decoding delays.

An instance of $\S$ is denoted by $S_U$ where $U$ is its cardinality. Its average packet decoding delay is denoted by $D_U$ and can be calculated using \eqref{eq:didnc_def}, where $u_{k}$ is now the index of the first maximal coding set in $\S_U$ that contains $\p_k$. Therefore, for the purpose of calculation, $\p_k$ can be removed from all the subsequent coding sets. After applying such removal to all the original data packets, the intersection between any two coding sets in $\S_U$ becomes empty. These coding sets are not necessarily maximal and we denote them by $\{\M^*_u\}$ to distinguish them from maximal ones. Below is an example.
\begin{Example}
Consider an instance $\S_3=\{\{\p_1,\p_2,\p_4\},\{\p_2$, $\p_5\},\{\p_3,\p_6\}\}$. After packet removal, the instance becomes: $\S_3=\{\{\p_1,\p_2,\p_4\},\{\p_5\},\{\p_3,\p_6\}\}$.
\end{Example}

Let us denote by $T_c(u)$ the number of targeted receivers of coding set $\M^*_u$. Without loss of generality we assume that:
\vspace{-5pt}
\begin{equation}\label{eq:Tc_order}
T_c(1)\geqslant T_c(2)\geqslant \cdots \geqslant T_c(U)
\end{equation}
It holds that $\sum_{u=1}^UT_c(u)=\sum_{k=1}^KT_k$. Then, as a variation of (\ref{eq:didnc_def}), the average packet decoding delay under $\S_U$ can also be calculated as:
\vspace{-1em}
\begin{equation}
D_U=\frac{1}{\sum_{u=1}^UT_c(u)}\sum_{u=1}^UuT_c(u)\label{eq:D_definition_new}
\end{equation}

The above two equations indicate the condition that the best/worst possible instances of $\S$ should satisfy:
\begin{enumerate}[C1.]
\item Because $T_c(u)>0$ for all $u$, $D_U$ is minimized if $\S_U$ has $T_c(1)=\sum_{k=1}^{K}T_k-(U-1)$ and $T_c(u)=1$ for $u\in[2,U]$. Since it is rare to have coding sets wanted by only one receiver, we relax this condition as $T_c(1)\gg T_c(2)\gg\cdots \gg T_c(U)$ and refer to such $\S_U$ as the best;
\item $D_U$ is maximized with a value of $(U+1)/2$ if $\S_U$ has $T_c(1)=T_c(2)=\cdots=T_c(U)$ and thus is the worst.
\end{enumerate}

We now propose different instances of $\S$ and obtain lower/upper bounds on $\DIDNC$ with different tightness.
\vspace{-1em}
\subsection{Loose Bounds}
For any given SFM, without loss of generality we assume that its conflict matrix $\mC$ belongs to $\C(K,M_0)$. By employing the loose bounds on $\UIDNC$ for $\C(K,M_0)$, we can derive loose bounds on $\DIDNC$.

\subsubsection{Loose lower bound}
The smallest possible cardinality of the instance $S_U$ is equal to the loose lower bound on $\UIDNC$, that is, $U=U(K,M_0)^-$. Thus, $D_U$ is minimized if $\S_U$ has:
\begin{equation}\label{eq:loose_low}
T_c(u)=
\begin{cases}
\sum_{k=1}^{K-U+1}T_k, & u=1 \\
T_{K-U+u},&u\in[2,U]
\end{cases}
\end{equation}

By substituting the above $\{T_c(u)\}$ into  (\ref{eq:D_definition_new}), a loose lower bound on $\DIDNC$ is obtained.

\subsubsection{Loose upper bound}
The largest possible cardinality of the instance $S_U$ is equal to the loose upper bound on $\UIDNC$, that is, $U=U(K,M_0)^+$. Thus, $D_U$ is maximized with a value of $(U(K,M_0)^++1)/2$ if $\S_U$ has uniform $\{T_c(u)\}$, as discussed in C2 after \eqref{eq:D_definition_new}.

\subsection{Tight Bounds}\label{sec:delay_tight}

\subsubsection{Tight lower bound} In the derivation of the tight lower bound on $\UIDNC$, we find a size-$U^-$ clique in the complementary IDNC graph $\cG$. Denote the original data packets included in this clique by $\{\p_1\cdots \p_{U^-}\}$, and without loss of generality assume that $T_1\geqslant T_2 \geqslant \cdots \geqslant T_{U^-}$. These original data packets must be sent separately because they are not connected in $\G$, i.e., they conflict. In this case, the smallest decoding delay takes place when all the remaining $K-U^-$ original data packets can be coded together with $\p_1$ in the first coding set. The sequence $\{T_c(u)\}$ is:
\begin{equation}
T_c(u)=\begin{cases}
T_1+\sum_{k=U^-+1}^KT_k, & u=1\\
T_u, & u=[2,U^-]
\end{cases}
\end{equation}

By substituting the above $\{T_c(u)\}$ into (\ref{eq:D_definition_new}), a tight lower bound on the minimum packet decoding delay is obtained and is denoted by $D_{U^-}$.

\subsubsection{Tight upper bound}
We use the IDNC solution $S_{U^+}$ found using the operation $\f(\G)$ in \eqref{eq:upper_operation} as our instance and thus its average decoding delay is our tight upper bound $D_{U^+}$.
\begin{figure*}[t]
\centering
\includegraphics[width=0.7\linewidth]{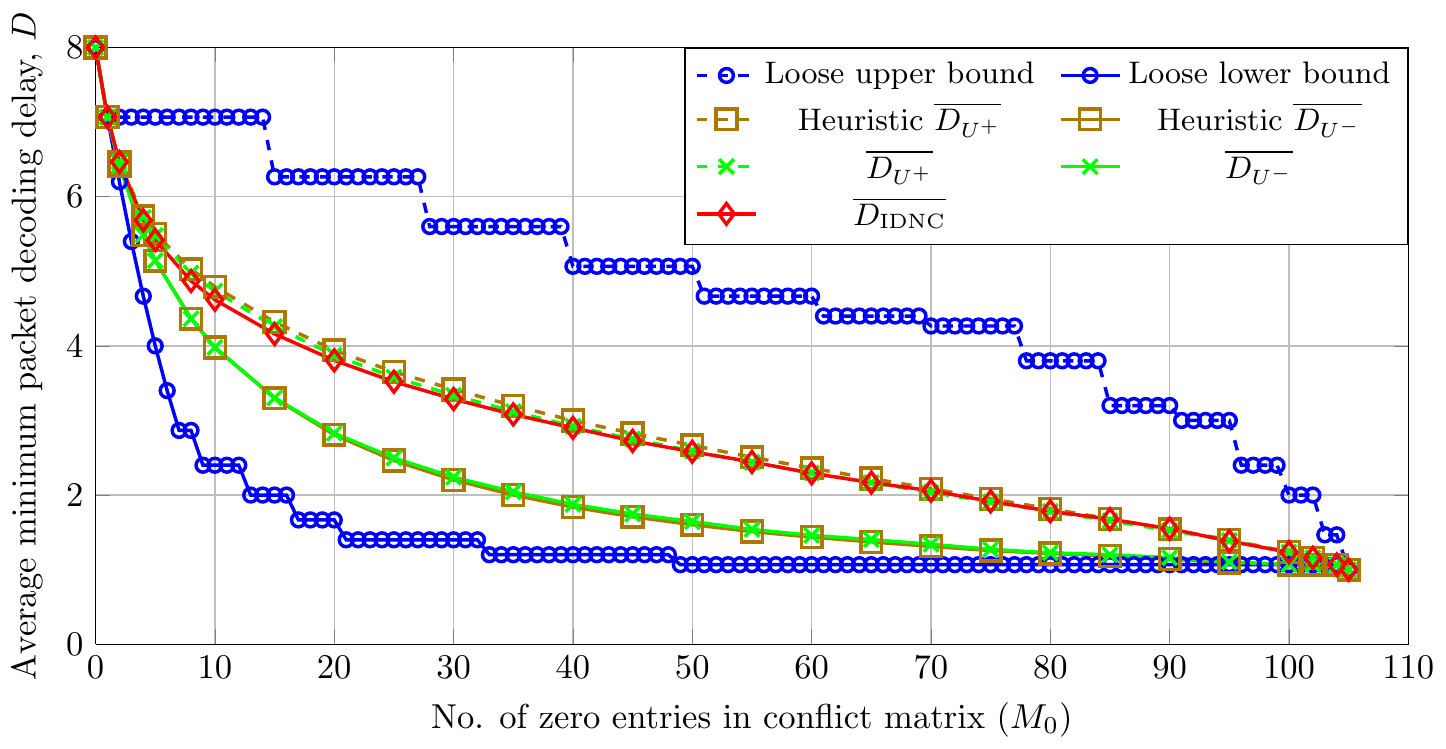}
\caption{Upper and lower bounds on the decoding delay with $K=15$ wanted original data packets.}
\label{fig:d_bounds_15}
\end{figure*}

\subsection{The Average Bounds on $\DIDNC$ and Their Tightness}\label{sec:delay_performance}

In this subsection, we obtain the average bounds on $\DIDNC$ using a similar method as for the average bounds on $\UIDNC$. For a given set of $\{K,M_0\}$, conflict matrices are randomly generated and the number of targeted receivers of the original data packets are also randomly generated. Their decoding delay bounds, as well as $\DIDNC$ under the optimal semi-online solution are calculated and averaged.

Simulation results for $K=15$ and $M_0\in[0,105]$ are plotted in \figref{fig:d_bounds_15}. The profiles of the average decoding delay bounds are similar to the average throughput bounds. The average loose bounds decrease with increasing $M_0$ in a staircase way, while the average tight bounds decrease gradually as $\overline{\DIDNC}$.

The main difference is that $\overline{\DIDNC}$ is much closer to $D_{U^+}$ than in the throughput case, and for $M_0>50$, the gap becomes negligible. The reason is that the IDNC solution $\S_{U^+}$ can be viewed as a greedy IDNC solution in terms of decoding delay. It transmits the largest maximal coding set first, which is likely to target the most receivers. This result, together with the small gap between $\overline{U^+}$ and $\overline{\UIDNC}$, indicate that $\f(\G)$ could be modified into a good heuristic IDNC coding algorithm, which will be discussed in the next section.

\section{Implementations}\label{sec:implementations}
In this section, we present the algorithmic implementations of IDNC. We first propose the optimal semi-online and fully-online IDNC coding algorithms and then their heuristic alternatives. We also employ a heuristic clique-finding algorithm to obtain heuristic tight bounds on the throughput and decoding delay performance of IDNC.

\subsection{Optimal IDNC Coding Algorithms}\label{sec:optimal_algorithms}

Our optimal semi-online IDNC coding algorithm finds the minimum collections of the conflict matrix in two steps:
\begin{enumerate}[{Step}-1]
\item \emph{Find all the maximal coding sets (cliques):}
This problem is NP-complete but has an efficient recursive algorithm called Bron-Kerbosch algorithm \cite{Bron_algorithm}. The group of all the maximal coding sets is denoted by $\mathcal{A}$.
\item \emph{Find minimum collections from $\mathcal{A}$:}
We propose an iterative algorithm in Algorithm \ref{algorithm:branch} to achieve it. The intuition behind this algorithm is that, if an original data packet belongs to $d$ maximal coding sets  in $\mathcal{A}$, one of these $d$ maximal coding sets has to be transmitted. In the extreme case of $d=1$, this maximal coding set must be sent. Below is an example of Algorithm \ref{algorithm:branch}.
\end{enumerate}

\begin{figure}
\centering
\includegraphics[width=0.9\linewidth]{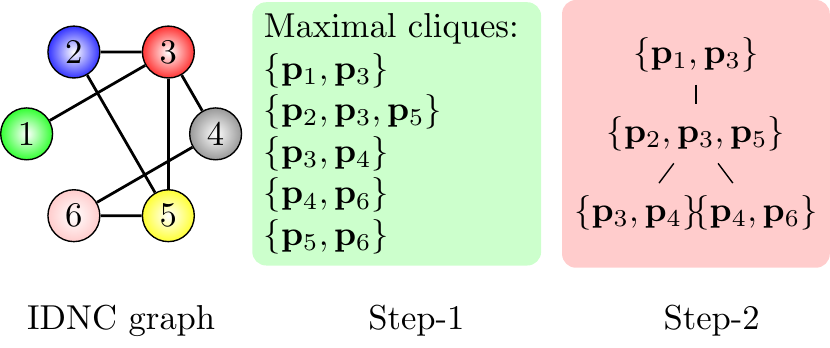}
\caption{An example of the proposed optimal IDNC algorithm.}\label{fig:min_collection}
\end{figure}

\begin{Example}

Consider the graph model in \figref{fig:min_collection}. In Step-1, we find all the maximal cliques: $\{\{\p_1,\p_3\}$, $\{\p_2,\p_3,\p_5\}$, $\{\p_3,\p_4\}$, $\{\p_4,\p_6\}$, $\{\p_5,\p_6\}\}$. Then in Step-2:
\begin{enumerate}
\item Since $\S$ is empty, none of the original data packets are included. Among them, $\p_1$ has a diversity of only one under $\overline{\S}$. Thus in the first iteration, the updated solution is $\S$$=\{\p_1,\p_3\}$;
\item The remaining original data packets are $\p_2,\p_4,\p_5,\p_6$. Among them $\p_2$ has a diversity of one under $\overline{\S}$. Thus in the second iteration, the updated solution is $\S=\{\{\p_1,\p_3\}$,$\{\p_2,\p_3,\p_5\}\}$;
\item The remaining original data packets are $\p_4$ and $\p_6$. They both have a diversity of two under $\overline{\S}$. We pick $\p_4$ and then branch: $\S_1=\{\{\p_1,\p_3\}$,$\{\p_2,\p_3,\p_5\}$,$\{\p_4,\p_5\}\}$ and $\S_2=\{\{\p_1,\p_3\}$,$\{\p_2,\p_3,\p_5\}$,$\{\p_4,\p_6\}\}$. Since $\S_2$ satisfies the diversity constraint, the algorithm ceases and returns $\S_2$ as the minimum collection.
\end{enumerate}
\end{Example}

If the above two-step coding algorithm outputs several minimum collections, different criterion can be used for selection, such as the smallest average packet decoding delay and the highest average packet diversity, etc. In our simulations, we select the one having larger diversities for data packets wanted by more receivers, i.e., the collection $\S$ that maximizes $\sum_{k=1}^{K}d_k T_k$ will be chosen, where $d_k$ is the diversity of $\p_k$ within $\S$ and $T_k$ is the number of targeted receivers of $\p_k$.

If fully-online feedback is allowed and computational cost at the sender is not an issue, the optimal fully-online IDNC scheme can be applied, where in every time slot, the sender calculates the optimal semi-online solution as above, but sends only the first maximal coding set and then collects feedback.

\begin{algorithm}[t]
\caption{Optimal minimum collections search}\label{algorithm:branch}
\begin{algorithmic}[1]
\STATE \textbf{input:} the group of all maximal coding sets, $\mathcal{A}$;
\STATE \textbf{initialization:} a set of collections $\mathcal{B}$ which only contains an empty collection, an iteration counter $u=1$;
\WHILE {no collection in $\mathcal{B}$ satisfies the diversity constraint,}
\WHILE {there is a collection in $\mathcal{B}$ with size $u-1$,}
\STATE Denote this collection by $\S=\{\M_1,\cdots,\M_{u-1}\}$. Denote the original data packets included in $\S$ by $\P=\bigcup_{i=1}^{u-1}\{\M_i\}$ and all the remaining original data packets by $\overline{\P}=\P_K\setminus\P$. Also denote the maximal coding sets excluded in $\S$ by $\overline{\S}=\mathcal{A}\setminus\S$;
\STATE Pick an original data packet, say $\p$, in $\overline{\P}$ which has the smallest diversity $d$ within $\overline{\S}$. Denote the $d$ coding sets which contain $\p$ by $\M'_1,\cdots,\M'_d$;
\STATE Branch $\S$ into $d$ new collections, $\S'_1,\cdots,\S'_d$. Then, add $\M'_1,\cdots,\M'_d$ to these collections, respectively. The size of the new collections are $u$;
\ENDWHILE
\STATE $u=u+1$;
\ENDWHILE
\STATE Output the collections in $\mathcal{B}$ that satisfy the diversity constraint.
\end{algorithmic}
\end{algorithm}

\subsection{Hybrid IDNC Coding Algorithms}\label{sec:hybrid_algorithms}
Algorithm \ref{algorithm:branch} is optimal because it finds all the possible minimum collections. However, it is also memory demanding because the number of candidature solutions usually grows exponentially after the branching in every iteration. Thus in this subsection, we propose a simple greedy alternative to it. We first choose the largest maximal coding set in $\mathcal{A}$. Then for the remaining original data packets that have not been covered, we look for a maximal coding set which comprises most of them. This iterative algorithm only produces one collection, which may be suboptimal because its cardinality may be greater than $\UIDNC$. The optimal clique finding in Step-1, together with this heuristic algorithm in Step-2, is referred to as the hybrid semi-online IDNC coding algorithm.

If fully-online IDNC is applied, after finding $\mathcal{A}$ in Step-1, we can greedily choose the maximal coding set in $\mathcal{A}$ that targets the maximum number of receivers. This algorithm is referred to as the hybrid fully-online IDNC coding algorithm.

To reduce the computational load due to Step-1, we resort to a fully-heuristic clique-finding algorithm next.

\subsection{Heuristic IDNC coding Algorithms}
A simple algorithm that heuristically finds the maximum (the largest maximal) clique in a graph is provided in Algorithm \ref{algorithm:clique}. The intuition behind this algorithm is that, a vertex is very likely to be in the maximum clique if this vertex has the largest number of edges incident to it. This algorithm has been employed in \cite{Rozner_Heuristic_clique,sorour:valaee:2010,sameh:valaee:globecom:2010} for fully-online IDNC. So we also refer to it as the heuristic fully-online IDNC coding algorithm. However, it has not been applied to semi-online IDNC and its computational complexity has not been identified yet.

\begin{algorithm}
\caption{Heuristic maximum clique search}\label{algorithm:clique}
\begin{algorithmic}[1]
\STATE \textbf{input}: graph $\G(\V,\E)$;
\STATE \textbf{initialization}: an empty vertex set $\V_{\textrm{keep}}$;
\WHILE {$\G$ is not empty}
\STATE Weight every vertex in $\G$ with the number of edges incident to it;
\STATE Find the vertex $v$ which has the largest weight;
\STATE Add $v$ to $\V_{\textrm{keep}}$;
\STATE Update $\G$ by deleting the vertices not connected to $v$, as well as deleting the edges incident to these vertices (\textit{Since these vertices cannot be part of the target clique, they can be ignored});
\STATE Update $\G$ by deleting $v$ and the edges incident to $v$ (\textit{since $v$ is already in $\V_{\textrm{keep}}$, there is no need to consider it anymore});
\ENDWHILE
\STATE Vertices in $\V_{\textrm{keep}}$ are all connected and thus form a clique.
\end{algorithmic}
\end{algorithm}

The computational complexity of this algorithm is polynomial in the number of original data packets $K$. The highest computational complexity occurs when the input graph is complete, i.e., all the vertices are connected to each other.
Under this scenario, only one vertex could be removed in each iteration (in Step 8) and thus, the size of the graph in the $i$-th iteration, $i\in[0,K-1]$, will be $K-i$. As a result of this, the highest computational complexity is $\sum_{i=0}^{K-1}K-i=K(K-1)/2$.  In practice, the graph size will shrink much faster after each iteration, and the number of iterations is usually smaller than $K$. Hence, the computational complexity of this algorithm is loosely upper-bounded by $K^2/2-K/2$. In other words, the computational complexity of this algorithm is $O(K^2)$.

\subsubsection{Heuristic bounds}
Here, we apply Algorithm \ref{algorithm:clique} to heuristically find the proposed tight bounds on the throughput, and then the corresponding tight bounds on the decoding delay. The results are shown in Figs. \ref{fig:u_bounds_15} and \ref{fig:d_bounds_15}, respectively. It is observed that the performance degradation due to the heuristic algorithm is marginal for both throughput and decoding delay. Therefore, the heuristic tight bounds could serve as reliable and efficient estimates of the throughput and decoding delay.

\subsubsection{Heuristic semi-online IDNC coding algorithm}
The operation $\f(\G)$ in \eqref{eq:upper_operation} can be implemented by using Algorithm \ref{algorithm:clique}. The outcome is a heuristic semi-online IDNC solution $\S_{U^+}$, which offers good throughput and decoding delay performance in the erasure-free scenario. However, since its cliques are disjoint, all the original data packets have a diversity of only one and thus are vulnerable to packet erasures in real systems.

To overcome this drawback, we propose a heuristic semi-online IDNC coding algorithm in Algorithm \ref{algorithm:encoding}, which is an extension of $\f(\G)$. The key idea here is that, in the $i$-th iteration, after finding clique $\M_i$, we try to enlarge this clique by adding previously covered vertices to it whenever possible, i.e., the vertices in $\V_{\textrm{covered}}=\bigcup_{j=0}^{i-1} \M_j$. By doing so, the diversity of the newly added vertices (packets) is increased by one. Below is an example:
\begin{Example}
Consider the graph $\G$ in \figref{fig:graph_example}. In the first two iterations, the algorithm will choose $\M_1=\{\p_1,\p_2,\p_4\}$ and $\M_2=\{\p_3,\p_6\}$, respectively, without any adding. In the third iteration, we have $\V_\textrm{covered}=\{\p_1,\p_2,\p_3,\p_4,\p_6\}$ and the algorithm can only choose $\M_3=\{\p_5\}$. Among all the original data packets in $\V_\textrm{covered}$, $\p_2$ can be added to $\M_3$. Thus $\M_3=\{\p_2,\p_5\}$. The process is then completed.
\end{Example}
\begin{table}
    \centering
    \begin{tabular}{|m{0.205\linewidth}|m{0.13\linewidth}|m{0.6\linewidth}|}
    \hline
   \bfseries{Feedback} & \bfseries{Optimality~~~~} & \bfseries{How to find the maximal coding sets in the solution}\\ \hline

    \multirow{3}{7.5em} {\textbf{semi-online}\\ (send whole solution then collect feedback)} & optimal & Run Bron-Kerbosch \cite{Bron_algorithm} in Step-1. Run Algorithm 1 in Step-2.\\ \cline{2-3}
    & hybrid & Run Bron-Kerbosch [28] in Step-1. Run greedy algorithm in Step-2 according to \secref{sec:hybrid_algorithms}.\\ \cline{2-3}
   & heuristic & Run Algorithm 3. \\ \hline
   \multirow{2}{7.5em}{\vspace{-18pt}   \\ \textbf{fully-online} \\ (send one coding set then collect feedback)} & optimal, hybrid & The same as the corresponding semi-online one, but only choose the maximal coding set that targets the most receivers.\\\cline{2-3}
   & heuristic & Run Algorithm 2.\\\hline
   \end{tabular}
\caption{A summary of the proposed IDNC schemes, \lowercase {where Step-1 and 2 refer to those described in Section VI-A}}
\label{tab:schemes_algorithms}
\end{table}
\begin{algorithm}
\caption{Heuristic semi-online IDNC algorithm}\label{algorithm:encoding}
\begin{algorithmic}[1]
\STATE \textbf{input}: a graph $\G(\V,\E)$;
\STATE \textbf{initialization}: generate an empty vertex set $\V_{\textrm{covered}}$, a working graph $\G_w=\G$, and a counter $i=0$;
\WHILE {$\V_{\textrm{covered}}\neq\V$}
\STATE Find the maximum clique in $\G_w$ using Algorithm \ref{algorithm:clique}. Denote it by $\M_i$ ;
\STATE Find the vertices in $\V_{\textrm{covered}}$ which are connected to $\M_i$. Denote their set by $\V_i$ (\textit{They are the candidate vertices that could be added to $\M_i$}.);
\STATE Generate a subgraph of $\G$ whose vertex set is $\V_i$. Denoted this subgraph by $\G'_i(\V_i,\E_i)$;
\STATE Find the maximum clique in $\G'_i$ using Algorithm 1, denoted it by $\M_i'$ (\textit{All vertices in $\M_i'$ are connected to each other and thus can all be added to $\M_i$.});
\STATE Update $\V_{\textrm{covered}}$ by adding vertices in $\M_i$ into it;
\STATE Update $\G_w$ by removing $\M_i$ from it;
\STATE Update $\M_i$ as $\M_i=\M_i\cup\M_i'$ (\textit{The new clique is at least as large as the old one and thus, provides higher packet diversity});
\STATE $i=i+1$;
\ENDWHILE
\end{algorithmic}
\end{algorithm}

\section{Simulations}\label{sec:simulations}
In this section, we numerically evaluate the throughput and
decoding delay performance of the proposed six IDNC
schemes with different combinations of feedback frequencies and algorithms, as shown in Table \ref{tab:schemes_algorithms}. Instead of $\UIDNC$ and $\DIDNC$, which indicate the best possible
performance of IDNC in terms of the throughput and packet decoding delay, we measure the total number of coded transmissions and the average packet decoding delay with
the presence of packet erasures. The total number of data
packets is $K_T=15$, the total number of receivers is
$N_T\in[5,45]$. Two simulations are carried out. In the
first simulation a small packet erasure probability of
$P_e=0.05$ is applied, while in the second simulation $P_e=0.2$. The throughput and decoding delay performance of RLNC are also plotted as references.

\begin{figure*}
\centering
\subfigure[Throughput]{\includegraphics[width=0.45\linewidth]{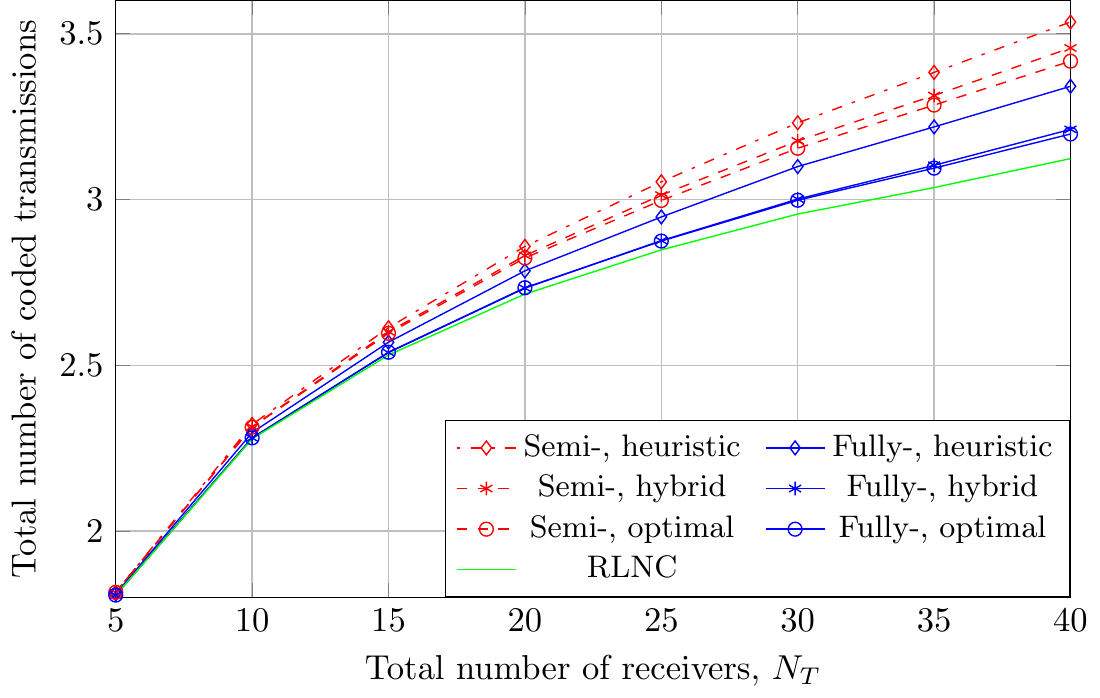}\label{fig:u_small_pe}}
\subfigure[Decoding delay]{\includegraphics[width=0.45\linewidth]{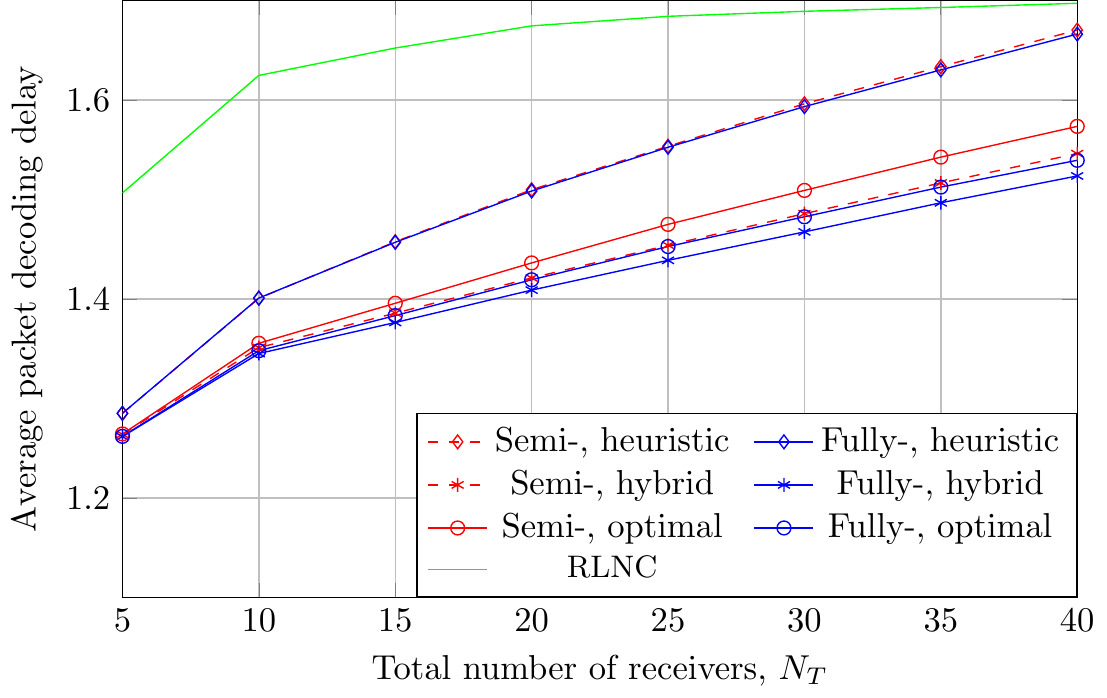}\label{fig:d_small_pe}}
\caption{Performance with a small packet erasure probability of $P_e=0.05$.}
\label{fig:performance_small_Pe}
\end{figure*}
\begin{figure*}
\centering
\subfigure[Throughput]{\includegraphics[width=0.45\linewidth]{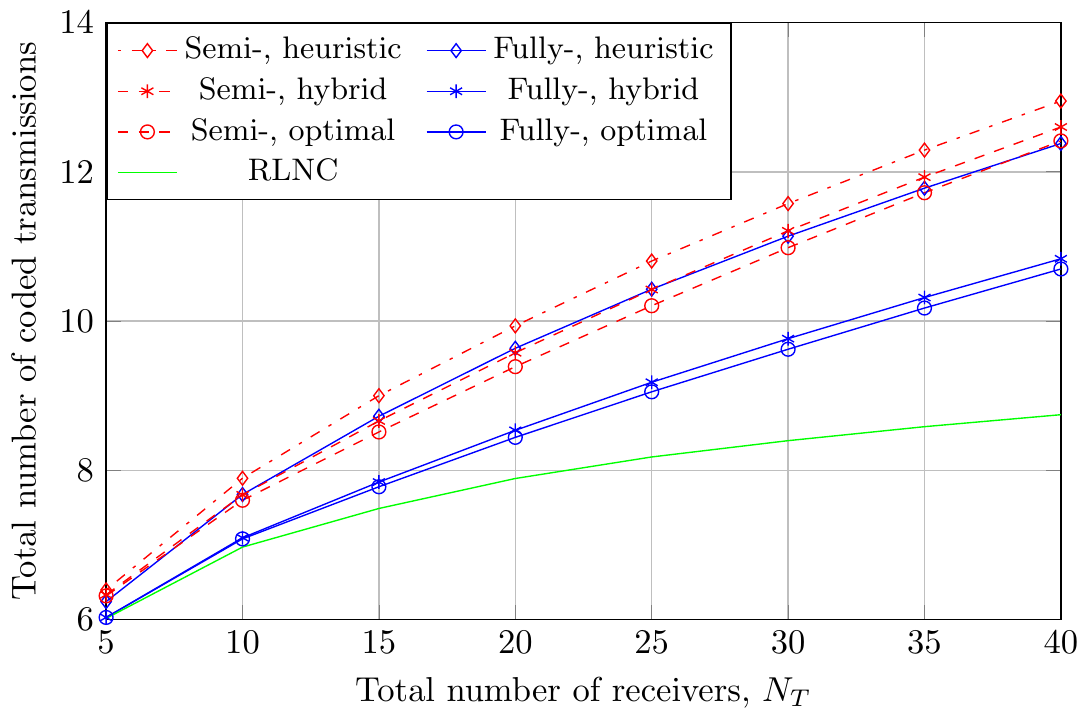}\label{fig:u_medium_pe}}
\subfigure[Decoding delay]{\includegraphics[width=0.45\linewidth]{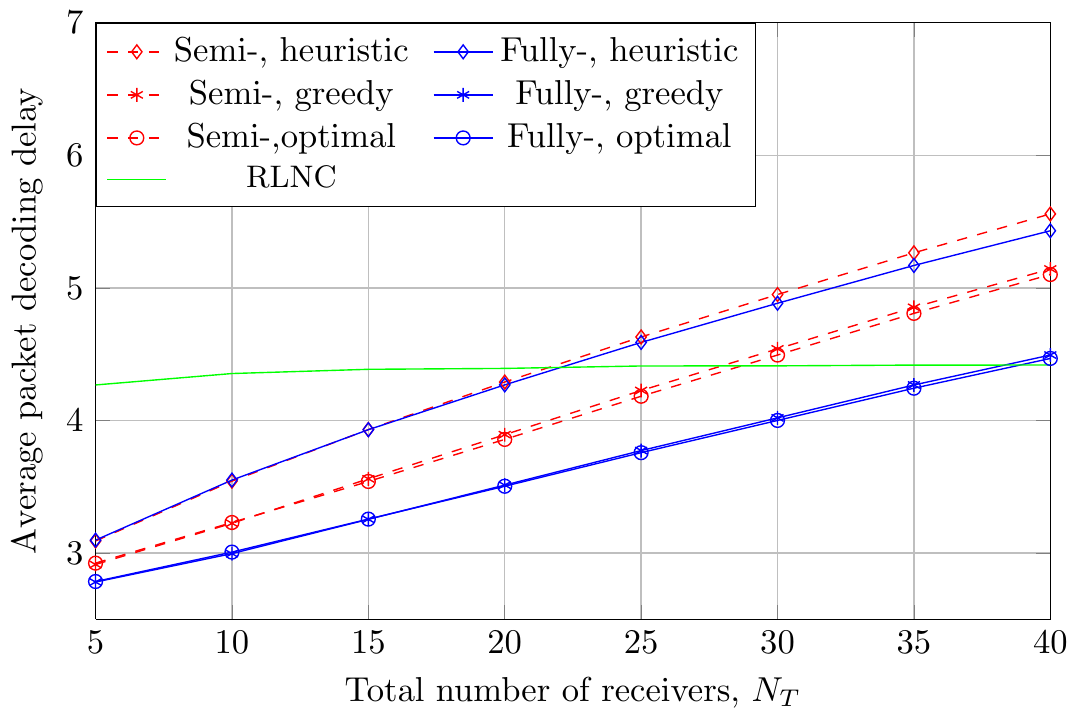}\label{fig:d_medium_pe}}
\caption{Performance with a medium packet erasure probability of $P_e=0.2$.}
\label{fig:performance_medium_Pe}
\end{figure*}

The simulation results are shown in Figs. \ref{fig:performance_small_Pe} and \ref{fig:performance_medium_Pe}. From these figures, we obtain some useful observations that could serve as simple guidelines for the implementation of IDNC:
\begin{enumerate}
\item The gaps between the semi-online and the corresponding fully-online schemes become larger when $P_e$ increases from 0.05 to 0.2. This result matches our expectation described in \secref{sec:optimal_schemes};
\item The gaps between the hybrid and the corresponding optimal schemes are marginal regardless of $P_e$ and $N_T$. The hybrid ones even exceed their (throughput) optimal counterparts in decoding delay performance when $P_e$ is small. See also remarks after \eqref{eq:didnc_def}. Hence, hybrid schemes could be preferable in practice, since they provide a good tradeoff between performance and computational load. This result also motivates the problem of finding decoding delay optimal IDNC solutions in future research.
\item The gaps between the heuristic semi-online and heuristic fully-online schemes are relatively small for all $P_e$ and $N_T$, especially in terms of decoding delay. Thus when the sender can only afford low computational load and it is primarily concerned with the decoding delay, applying semi-online feedback frequency is sufficient;
\end{enumerate}

A cross comparison between the optimal fully-online IDNC and RLNC shows that, in terms of throughput, the performance of IDNC is close to RLNC for all values of $N_T$ when $P_e=0.05$, but the gap increases with $N_T$ when $P_e=0.2$. In terms of decoding delay, the performance of IDNC is always much better than RLNC when $P_e=0.05$. Such superiority becomes marginal with the increase of  $N_T$ and $P_e$. We conclude that, first, the performance of IDNC is more vulnerable to the increase in the number of receivers and bad channel quality than RLNC. Second, there is no clear winner between them when we consider both throughput and delay.

\section{Conclusion and Open Questions}
In this paper, we presented a systematic study of instantly decodable network coding (IDNC) in the single-hop wireless broadcast scenario.  We obtained the optimal solution of IDNC in terms of the completion time (as a measure of throughput) in the fully-online system where receivers' feedback about the status of their received and lost packets are available at the sender after every single IDNC transmission. We also studied a semi-online IDNC scheme where receivers' feedback is available after a number of IDNC transmissions, and correspondingly proposed the optimal semi-online IDNC solution. However, since finding these optimal solutions are computationally complex, efficient heuristic algorithms were also proposed. Moreover, our studies on the optimal IDNC solution resulted in finding useful loose and tight upper- and lower-bounds on the best throughput and decoding delay performance of IDNC. While the loose bounds reveal the performance limits of IDNC, the tight bounds provide efficient estimates of the performance of IDNC. Extensive simulations were carried out to evaluate the throughput and decoding delay performance of the proposed schemes. The simulation results can guide simple implementations of IDNC.

There are also many interesting directions that this work could be extended to. For example,  the optimal decoding delay performance of IDNC is yet widely unaddressed in the literature mainly due to 1) the lack of a commonly accepted definition of decoding delay; and 2) the complicated interplay of decoding delay and throughput. Another interesting direction is to extend this work to the general IDNC scheme, or allow receivers to store non-instantly decodable packets.

\appendices
\renewcommand\thesection{Appendix \Alph{section}}
\section{Proof of Theorem \ref{theo:chromatic}}\label{app:1}

We prove that if there is a collection $\S=\{\M_1,\cdots,\M_U\}$ satisfying the diversity constraint, there is also a $U$-coloring solution of~$\cG$ and \emph{vice versa}. An eligible $U$-coloring solution is denoted by $\{\K_1,\cdots,\K_U\}$ and has the following three properties: 1) the vertices in the same set $\K_i$ share the same color; 2) any two of vertices in the same $\K_i$ are not connected; 3) the intersection between any two sets $\K_i$ and $\K_j$ is empty.

If $\S=\{M_1,\cdots,\M_U\}$ satisfies the diversity constraint, we can always construct a new group of vertex sets as follows, which also satisfy the diversity constraint:
Set $\K_1=\M_1$, then sequentially $\K_2=\M_2\setminus\K_1$, $\K_3=\M_3\setminus\bigcup\{\K_1,\K_2\},\cdots$, $\K_U=\M_U\setminus\bigcup_{i=1}^{U-1}\{\K_i\}$. Following this construction process, it is clear that the intersection between any two sets $\K_i$ and $\K_j$ is empty, thus property 3) is satisfied. Since $\K_i\subseteq \M_i$, every vertex in the same $\K_i$ is connected to each other under $\G$ and therefore, they are all disconnected under $\cG$ and property 2) is satisfied. Hence, if we assign $U$ colors to $K_i,\cdots,\K_U$, they form a $U$-coloring solution of $\cG$. On the other hand, a $U$-coloring solution of $\cG$ is also a valid IDNC collection in which all the data packets have a diversity of one. Hence, the minimum number of colors, $\chi(\cG)$, is also the minimum collection size of the corresponding conflict matrix.

\section{Proof of Theorem \ref{theo:chromatic_minus}}\label{app:2}

Suppose $\M$ is a maximal clique of $\G$ and $\chi(\cG)=U$. Assuming the chromatic number of $\cGn=\cG\setminus\M$ is $U'<U-1$, there exists at least one $U'$-coloring solution of $\cGn$, denoted by $\{\K_1,\cdots,\K_{U'}\}$. If this is the case, $\{\K_1,\cdots,K_{U'},\M\}$ is a eligible $(U'+1)$-coloring of $\cG$ and $U'+1<U$, contradicting the assumption that $\chi(\cG)=U$. Thus $\chi(\cGn)$ is at least $U-1$.

The chromatic number of $\cGn$ will still be $U$ when the removed $\M$ does not belong to any minimum collection of $\mC$. If $\chi(\cGn)=U-1$, there exists at least one $(U-1)$-coloring solution of $\cGn$, denoted by $\{\K_1,\cdots,\K_{U-1}\}$. Then $\{\K_1,\cdots,\K_{U-1},\M\}$ is a valid minimum collection, contradicting that $\M$ does not belong to any minimum collection of $\mC$. Thus $\chi(\cGn)=U$ in this case.

\end{document}